\documentclass[times,twocolumn,final]{elsarticle}
\usepackage[margin=0.7in]{geometry}
\usepackage{framed,multirow}
\usepackage{booktabs}
\usepackage{tabularray}
\UseTblrLibrary{booktabs}
\usepackage{caption}
\usepackage{boldline} 
\usepackage{fontawesome5}
\usepackage{amssymb,amsmath}
\usepackage{romannum}
\AtBeginDocument{\pagenumbering{arabic}}
\usepackage{pifont}

\usepackage{latexsym}
\usepackage{mathrsfs} 
\usepackage{soul}
\usepackage{url}
\usepackage[dvipsnames,table,xcdraw, svgnames,x11names]{xcolor}
\usepackage{makecell}
\usepackage{xcolor}

\usepackage{blindtext}
\usepackage{graphicx}
\usepackage{multirow}

\usepackage[listings,skins,breakable]{tcolorbox}
\usepackage[colorlinks]{hyperref}
\usepackage[printonlyused,withpage]{acronym}
\usepackage[nameinlink]{cleveref}
\usepackage{caption}
\usepackage{subcaption}
\newcolumntype{P}[1]{>{\raggedright\arraybackslash}p{#1}}

\definecolor{newcolor}{rgb}{.8,.349,.1}
\definecolor{maroon}{cmyk}{0,0.87,0.68,0.32}
\definecolor{lightorange}{rgb}{1,0.753,0.478}
\usepackage{array}
\newcolumntype{L}[1]{>{\raggedright\arraybackslash}m{#1}}
\newcolumntype{M}[1]{>{\raggedright\arraybackslash}m{#1}}
\newcommand{\cell}[1]{\begin{minipage}[t]{\linewidth}\raggedright #1\end{minipage}}
\newcommand{\tabitem}{\textbullet\ }

\begin{document}

\begin{frontmatter}

\title{Harmonization in Magnetic Resonance Imaging: A Survey of Acquisition, Image-level, and Feature-level Methods}%

\author{Qinqin Yang\corref{cor1}\textsuperscript{a}}
\author{Firoozeh Shomal-Zadeh\textsuperscript{b}, Ali Gholipour\textsuperscript{a,c}}

\cortext[cor1]{Corresponding author: Qinqin Yang, $\:$%
  Department of Radiological Sciences, University of California Irvine, Irvine, 856 Health Sciences Quad, Irvine, CA, 92697,$\:$%
  E-mail: qinqin.yang@uci.edu.}

\address[1]{Department of Radiological Sciences, University of California Irvine, Irvine, CA 92697, USA}
\address[2]{Department of Radiology, University Hospitals Cleveland Medical Center/Case Western Reserve University, Cleveland, OH 44106, USA}
\address[3]{Department of Electrical Engineering and Computer Science, University of California Irvine, Irvine, CA 92697, USA}

\begin{abstract}Magnetic resonance imaging (MRI) has greatly advanced neuroscience research and clinical diagnostics. However, imaging data collected across different scanners, acquisition protocols, or imaging sites often exhibit substantial heterogeneity, known as “batch effects” or “site effects.” These non-biological sources of variability can obscure true biological signals, reduce reproducibility and statistical power, and severely impair the generalizability of learning-based models across datasets. Image harmonization is grounded in the central hypothesis that site-related biases can be eliminated or mitigated while preserving meaningful biological information, thereby improving data comparability and consistency. This review provides a comprehensive overview of key concepts, methodological advances, publicly available datasets, and evaluation metrics in the field of MRI harmonization. We systematically cover the full imaging pipeline and categorize harmonization approaches into prospective acquisition and reconstruction, retrospective image-level and feature-level methods, and traveling-subject-based techniques. By synthesizing existing methods and evidence, we revisit the central hypothesis of image harmonization and show that, although site invariance can be achieved with current techniques, further evaluation is required to verify the preservation of biological information. To this end, we summarize the remaining challenges and highlight key directions for future research, including the need for standardized validation benchmarks, improved evaluation strategies, and tighter integration of harmonization methods across the imaging pipeline.

\end{abstract}

\end{frontmatter}

\section{Introduction} \label{sec:intro}
Magnetic resonance imaging (MRI) has had a profound impact on medicine, with widespread applications in medical and neuroscience research, computer-aided diagnosis, longitudinal monitoring, and image-guided interventions. To advance scientific discovery and bridge the gap between research and clinical practice, the collection and sharing of large-scale imaging datasets across sites has become increasingly essential \citep{RN134,RN133,RN121,RN136,RN127,RN132,RNintro02,RN143}. Multi-center studies that aggregate large and diverse samples not only enhance statistical power, particularly important for investigating rare or low-prevalence diseases, but also provide broader coverage of key biological variables such as age, sex, race, geographic location, socioeconomic status, and disease subtypes. The increased sample size and heterogeneity also improve the ability of studies to detect subtle yet meaningful effects in high-dimensional spaces of variables and confounders \citep{RN124,RN143}.

One of the major challenges in integrating multi-center, multi-site imaging data for joint analysis lies in non-biological technical variability. This variability, often referred to as scanner effects or batch effects, arises from differences in hardware and software across manufacturers, imaging sequences and protocols, as well as image processing pipelines and related techniques \citep{RN144,RN138,RN141,RN140}. This heterogeneity substantially compromises cross-site comparability and, consequently, degrades analytical performance, particularly in the era of deep learning \citep{RN145,RNSCIDATA_HARMML}.

To avoid the challenges associated with directly comparing heterogeneous imaging data, a conventional approach is meta-analysis, in which each site performs its analysis independently and the results are subsequently combined \citep{RNmeta1,RN149,RN148,RN146}. However, meta-analysis typically relies on group-level statistical and clinical summaries, making it difficult to perform detailed modeling or adjustments at the individual level. Furthermore, when participant distributions are imbalanced across sites, site-specific estimates may introduce systematic biases. In studies with limited imaging sample sizes, fluctuations in parameter estimation during procedures may further compromise the stability of statistical inference. In contrast, mega-analysis enables the joint analysis of all raw imaging data within a unified framework \citep{RNmega1,RNmega2,RNmega3,RNmega4}, thereby facilitating more comprehensive use of individual-level information. However, this strategy imposes more stringent requirements on data harmonization, as pooling datasets from different centers may amplify non-biological variability, particularly that arising from differences in imaging protocols. Therefore, effective harmonization, which aims to reduce non-biological variability while preserving. biologically meaningful information, is a critical prerequisite for enabling reliable mega-analyses \citep{RN135,RN39}.

Early MRI harmonization efforts primarily focused on standardizing acquisition protocols, together with digital operations on image intensities and histograms (e.g., normalization and histogram matching) \citep{RNintro01,RN51}. Subsequently, inspiration from genomics research on removing batch effects led to the development of traditional statistical model-based approaches (e.g., ComBat and its variants) operating on image-derived features \citep{RN77,RN99}. In recent years, deep learning has brought new momentum to MRI harmonization. On the one hand, powerful image synthesis architectures, such as generative adversarial networks (GANs) and diffusion models, have enabled a series of image-level harmonization methods \citep{RN56,RN61}. On the other hand, the strong feature extraction and nonlinear modeling capabilities have further driven the development of learning-based feature-level harmonization methods built upon traditional statistical models \citep{RN68,RN73,RN72}. Meanwhile, the role of MRI harmonization has expanded from supporting multi-site statistical analysis to facilitating learning-based downstream tasks, such as tissue segmentation, disease classification, and age prediction.

This review focuses on multi-site harmonization methods for MRI data, while the underlying principles are broadly applicable to other medical imaging modalities. MRI exhibits pronounced inter-site heterogeneity due to its inherent characteristics, including multiple field strengths, diverse imaging protocols and modalities, and a wide range of quantitative parameters. Several previous surveys have reviewed MRI harmonization. For example, Hu et al. primarily focused on retrospective approaches based on statistical models and early deep learning methods \citep{RN39}, whereas Pinto et al. and Abbasi et al. mainly concentrated on brain diffusion MRI and structural MRI, respectively \citep{RN43,RN42}. In contrast, our review extends prior work by providing a unified perspective on harmonization across the entire MRI pipeline, including prospective acquisition-level strategies (e.g., vendor-agnostic pulse sequences and harmonized reconstruction) as well as retrospective image- and feature-level methods (\Cref{fig:fig1}). In addition, we highlight recent advances in deep learning-based harmonization, including approaches that leverage multi-contrast priors, source-free methods, and emerging frameworks that integrate statistical modeling with deep learning to improve interpretability. Rather than providing an exhaustive survey, we focus on representative methodological developments that capture the core ideas of each category and aim to provide insights that may guide future research in this area.

\begin{figure*}[t]
 \centering
 \includegraphics[width=0.95\textwidth]{Figure-01.jpg}
 \caption{Overview of harmonization strategies across the entire MRI pipeline, covering image acquisition, reconstruction, post-processing, and feature-level analysis, and including representative methods discussed in this review. dMRI, diffusion-weighted MR imaging; T1WI, T1-weighted imaging; T2WI, T2-weighted imaging; RBV, regional brain volume; CT, cortical thickness; FA, fractional anisotropy; BART, Berkeley Advanced Reconstruction Toolbox; ISMRMRD, International Society for Magnetic Resonance in Medicine Raw Data; RISH, Rotationally Invariant Spherical Harmonic; TS, Traveling Subject; RAVEL, Removal of Artificial Voxel Effect by Linear regression; RELIEF, REmoval of Latent Inter-scanner Effects through Factorization; cVAE, conditional Variational Autoencoder.}
 \label{fig:fig1}
\end{figure*}

\section{Background}\label{sec:background}

\begin{table*}[!ht]
    \captionsetup{font=normalsize} 
    \centering
    \caption{Non-biological sources of variability in MRI data}
    \label{tab:tab1}
    \small
    \renewcommand{\arraystretch}{1.08}
    \begin{tabular}{P{0.22\textwidth} P{0.22\textwidth} P{0.52\textwidth}}
        \toprule
        \textbf{Source of variability} & \textbf{Category} & \textbf{Examples} \\
        \midrule

        \multirow[t]{6}{=}{\makecell[l]{Hardware-\\related factors}}
        & Scanner vendor & Siemens, GE, Philips, United Imaging \\
        & Field strength & 0.55T, 1.5T, 3T, 5T, 7T \\
        & Gradient system & Maximum gradient strength, slew rate, gradient nonlinearity \\
        & RF transmit system & Amplifier characteristics, parallel transmission (pTx) \\
        & RF receiver system & Coil geometry, number of channels, analog-to-digital converter \\
        & Shim system & $B_0$ and $B_1$ field homogeneity \\
        \midrule

        \multirow[t]{2}{=}{\makecell[l]{Acquisition-level\\differences}}
        & Pulse sequence design 
        & RF pulse design (slice profile), contrast mechanism (T1w, T2w, T2*w, diffusion), preparation modules (inversion recovery, MT, fat suppression); readout trajectory (single-shot, multi-shot, Cartesian, non-Cartesian), vendor-specific implementations \\
        & Imaging parameters or protocol
        & Contrast (TE, TR, TI, flip angle), spatial resolution (FOV, matrix size, slice thickness), signal-to-noise ratio (receiver bandwidth, NEX), imaging acceleration (parallel imaging, partial Fourier, SMS), diffusion encoding parameters (b-value, gradient directions) \\
        \midrule

        \multirow[t]{6}{=}{\makecell[l]{Post-processing\\effects}}
        & Reconstruction and coil combination 
        & SENSE, GRAPPA, compressed sensing, deep learning, sum-of-squares combination, adaptive combination \\
        & Image normalization and filtering 
        & Intensity normalization, denoising, raw filter (elliptical filter, Hamming filter), image filter (prescan normalize, $B_1$ correction, Gaussian filter) \\
        & Artifact correction
        & Odd-even phase correction, Gibbs-ringing artifact correction, distortion correction, N4 bias correction, motion correction \\
        & Feature extraction pipelines 
        & Segmentation, registration, texture operators, model fitting \\
        \bottomrule
    \end{tabular}

    \caption*{\footnotesize
    \textbf{Abbreviations:} MT, magnetization transfer; TE, echo time; TR, repetition time; TI, inversion time; FOV, field of view; NEX: number of excitations; SENSE, sensitivity encoding; GRAPPA, generalized autocalibrating partially parallel acquisitions; SMS, Simultaneous multislice.
    }
\end{table*}

\subsection{Objective of MRI Harmonization}
The primary objective of MRI harmonization is to enable the integration and joint analysis of multi-site and multi-batch imaging data by reducing non-biological variability while preserving biologically meaningful variation. Therefore, harmonization is not intended to recover an absolute “ground truth” or to achieve complete elimination of systematic biases, but rather to enhance the reliability of comparisons at both the individual and group levels. On this basis, it supports a wide range of applications, including mega-analysis, biomarker discovery, quantitative analysis, multi-site clinical studies, and learning-based downstream tasks.

The collection and analysis of MRI data involve a complex and variable set of procedures, including subject recruitment and selection, imaging hardware, protocol design, image reconstruction, and downstream analysis models. Variations at any of these stages, across imaging sites, subjects, or longitudinal scans, can introduce systematic differences into the final measurements. To achieve effective harmonization, it is essential to understand the major sources of variability in MRI data, which can be broadly categorized into biological and non-biological factors \citep{RNfactors,RN130,RNbg1,RNbg2,RNbg3}. The primary biological factors include age, sex, ethnicity, and disease status. These factors contribute to sampling bias in the data and can be incorporated as covariates in harmonization models for further modeling. In contrast, non-biological factors include imaging hardware, pulse sequences, acquisition protocols, and post-processing algorithms. These factors introduce measurement bias into the data and are summarized in \Cref{tab:tab1}.

\subsection{Problem formulation and methodology}\label{sec:bacif}

In general, the observed measurement $x_{ij}$ can be viewed as arising from both biologically meaningful factors $b_i$ and site-related factors $s_j$, together with residual noise $\epsilon$, i.e.,
\begin{equation}
x_{ij} = f(b_i, s_j) + \epsilon_{ij}
\end{equation}
where $i$ and $j$ index subjects (samples) and sites (or scanner settings), respectively. The goal of harmonization is to mitigate the influence of $s_j$ while preserving biologically meaningful variation associated with $b_i$. Under this general formulation, different harmonization paradigms operate at distinct stages of the imaging and analysis pipeline.

\subsubsection{Prospective approaches}
Prospective harmonization refers to strategies that are deliberately planned prior to data acquisition with the goal of minimizing anticipated sources of variability at the source. Based on Eq.~(1), this can be expressed as:
\begin{equation}
\bar{x}_{ij} \approx f(b_i, s^*) + \epsilon_{ij}
\end{equation}
where $s^*$ denotes a standardized acquisition setting. A common approach involves standardizing scanner models and acquisition protocols in advance to reduce differences introduced during image acquisition. Building on this foundation, this review highlights recent advances such as vendor-agnostic pulse sequences and harmonized image reconstruction methods, which aim to overcome traditional barriers to acquisition consistency through open-source and easily implementable solutions \citep{RN14,RN16,RN30,RN21}. These innovations further enhance the effectiveness of prospective harmonization. In addition, the use of traveling subjects represents another important prospective strategy, whereby the same individuals are scanned across multiple sites to obtain matched datasets \citep{RNTS1,RN83,RN118,RN153}. This design provides a valuable reference for quantifying and correcting systematic inter-site differences during analysis, thereby supporting the effective removal of non-biological variability.

\subsubsection{Retrospective approaches}
Retrospective harmonization refers to the application of harmonization techniques to existing, heterogeneous multi-site datasets after data acquisition. Due to the availability of large-scale public datasets and their cost-effectiveness and flexibility relative to prospective approaches, such methods currently dominate the field. Existing retrospective strategies span traditional image processing, statistical modeling, and deep learning-based approaches, and can be broadly categorized into image-level and feature-level methods. 

Image-level harmonization directly modifies voxel intensities, typically formulated as an image-to-image translation problem, aiming to standardize contrast, sharpness, and signal-to-noise ratio (SNR) across sites so that the resulting images appear as if acquired under comparable conditions \citep{RN36,RN51,RN66}. Based on Eq.~(1), image-level harmonization seeks a transformation $H$ applied directly to images and can be expressed as:
\begin{equation}
\bar{x}_{ij} = H(x_{ij}) \approx f(b_i) + \epsilon_{ij}
\end{equation}
where the harmonized image is expected to retain biological content while reducing site-related variation. These harmonized images can subsequently support a wide range of downstream analyses but also carry the risk of introducing artifacts or distorting anatomical structures, particularly when complex learning-based generative models are employed. In contrast, feature-level harmonization operates on derived image features (e.g., regional volumes, cortical thickness, functional connectivity, or radiomic features), enabling the use of statistical models that explicitly incorporate biological covariates to remove site effects \citep{RN94,RN46,RN99,RN5,RN3}. Similarly, this can be expressed as:
\begin{equation}
z_{ij} = \phi(x_{ij}), \quad \tilde{z}_{ij} = H(z_{ij}) \approx f(b_i) + \epsilon_{ij}
\end{equation}
where $\phi$ represents a feature extractor. This approach reduces the risk of altering image appearance and is computationally efficient, but it depends on the feature extraction pipeline and limits the reusability of the harmonized data.

\section{Harmonized Data Acquisition}\label{sec:acq}
An MRI system consists of two main components: hardware and software (\Cref{fig:fig1}). Although exact matching of both components would ideally enable optimal multi-site harmonization, this is rarely achievable in large-scale studies. Compared with hardware, harmonizing software offers greater flexibility and can be categorized into vendor-agnostic pulse sequence and harmonized image reconstruction approaches.

\subsection{Vendor-agnostic pulse sequence}
\begin{figure*}[t]
 \centering
 \includegraphics[width=\textwidth]{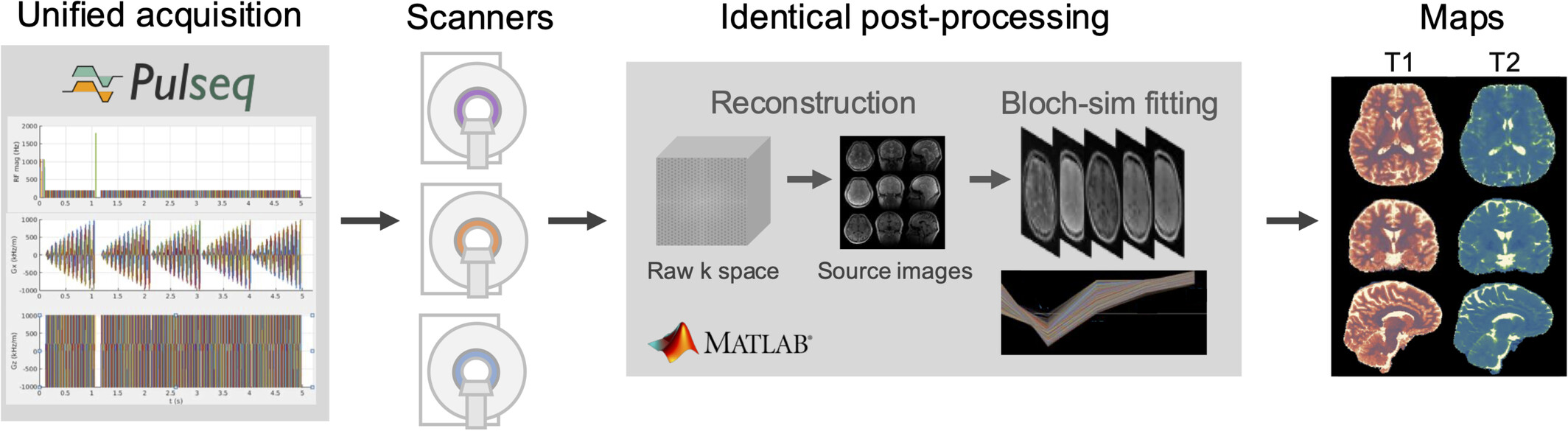}
 \caption{Harmonized acquisition and reconstruction workflow proposed in \citep{RN12}. The pulse sequence was implemented using Pulseq to ensure identical configurations across scanners and vendors. All post-processing steps including image reconstruction and quantitative parameter fitting were performed offline using a consistent pipeline.}
 \label{fig:fig2}
\end{figure*}

Since pulse sequences serve as the core of MR image formation, their harmonization offers an approach to addressing site effects at the source. However, due to differences in the underlying implementation of pulse sequences from different vendors, signal discrepancies may still arise even when identical acquisition parameters (e.g., TE, TR, FOV, matrix size) are used \citep{RN17}. These inconsistencies arise from differences in sequence implementation, including preparation modules, dephasing strategies (e.g., spoiler and crusher gradients), readout trajectories, and RF pulse shapes and profiles (\Cref{tab:tab1}), most of which are not accessible or adjustable through the user interface \citep{RN14, RN12}. To address this challenge and enhance consistency at sequence level, several vendor-agnostic or open-source pulse sequence platforms have been developed over the past decade, including Pulseq \citep{RN14}, gammaSTAR \citep{RN16} and RTHawk \citep{RN18}. For example, Pulseq enables modular pulse sequence programming in MATLAB and Python, allowing extensive and detailed control over RF pulses, gradient waveforms, and inter-module interval. The resulting sequence is compiled into a standardized .seq file, which can then be interpreted and executed by vendor-specific backends for MRI scanning (\Cref{fig:fig2}). Additionally, Pulseq can be integrated with various MRI simulation and graphical sequence design tools, further alleviating the steep learning curve of pulse sequence development \citep{RNMTRK}.

Recent studies provide empirical evidence supporting these approaches. Liu et al. \citep{RN9} systematically evaluated a single-shell diffusion MRI sequence implemented using Pulseq across two scanners from different vendors, using standard error as a metric of repeatability. For mean diffusivity in phantoms, the Pulseq sequence demonstrated 2.5-fold superior inter-scanner reproducibility compared to vendor-provided sequences. In human brain imaging, a Pulseq sequence reduced inter-scanner standard error in fractional anisotropy by 35–50\% across various brain regions. In addition to diffusion MRI, vendor-agnostic pulse sequence tools have also been validated in quantitative MRI (qMRI) applications, including chemical exchange saturation transfer \citep{RN11}, brain T1 and T2 mapping \citep{RN12,RN13}, and myocardial T1 mapping \citep{RN10,RN8}. In the work by Karakuzu et al. \citep{RN17}, combining RTHawk-based acquisition harmonization with a unified parameter quantification workflow led to statistically significant reductions in inter-vendor variability for T1, magnetization transfer ratio, and magnetization transfer saturation index measurements. Although current preliminary results are encouraging, it remains unclear to what extent vendor-agnostic or open-source pulse sequence tools can mitigate site effects, as they have yet to be widely implemented or validated at scale.

\subsection{Harmonized data reconstruction}
The raw MRI signal acquired from the scanner is one-dimensional complex-valued data. To generate the final image, this signal should be filled into a predefined k-space trajectory and then transformed via Fourier transformation, which is known as image reconstruction. Differences in reconstruction pipelines can introduce non-negligible variability, contributing to a lack of harmonization across sites or vendors. These differences may arise from multiple factors, including pre- and post-reconstruction distortion and phase correction, k-space gridding, partial Fourier reconstruction, multi-coil parallel reconstruction, and coil combination strategies (\Cref{tab:tab1}) \citep{RN22}. While vendors provide access and control over some of the options and parameters through user interfaces during acquisitions, a full control over the entire raw data processing pipeline often requires additional programming within the vendor software environment, which may not always be available or if available, may not be straightforward.

Offline open-source reconstruction toolboxes offer alternative, promising opportunities for standardizing the image reconstruction process. Representative examples include the Berkeley Advanced Reconstruction Toolbox (BART) \citep{RN30} and the Michigan Image Reconstruction Toolbox (MIRT) \citep{RN29}. For instance, BART not only implements conventional parallel imaging algorithms but also provides general-purpose solutions for non-Cartesian, model-based, and deep learning-based reconstruction \citep{RN27}. Its cross-platform, open-source, and multi-language support (Linux terminal, MATLAB, and Python) make it accessible and easy to integrate into diverse research workflows. Prior to reconstruction, inconsistencies in raw data formats across vendors pose a significant challenge. The ISMRMRD (International Society for Magnetic Resonance in Medicine Raw Data) \citep{RN24} framework addresses this issue by providing a standardized format that harmonizes vendor-specific raw data and headers, thereby facilitating consistent and reproducible reconstruction pipelines.

Despite their ease of use, offline open-source reconstruction toolboxes have inherent limitations that restrict their clinical scalability. These include the lack of real-time quality control and the requirement to store large raw datasets. To address these challenges, online reconstruction frameworks, e.g., Gadgetron \citep{RN21}  and FIRE (Framework for Image Reconstruction Environments) \citep{RNFIRE,RNFIRE2}, as well as cloud-based remote reconstruction systems, have been proposed. Gadgetron adopts a modular, streaming-based architecture and incorporates a wide range of extensible toolboxes, enabling real-time reconstruction through GPU or multithreaded CPU acceleration. It also supports advanced features such as automated motion tracking and scan planning, thereby minimizing heterogeneity arising from operator-dependent variability across sites. A notable example is the HERON framework, which leverages image-based real-time motion estimation to dynamically adjust fetal diffusion MRI (dMRI) acquisition, thereby mitigating the impact of unpredictable fetal motion \citep{RN20}. Similar strategies have also been extended to fetal functional MRI (fMRI), qMRI, and automated cardiac scan planning \citep{RN23,RN28,RN26,RNJMRI_PLANE,RNRADAI_PLANES}. Online reconstruction in these challenging applications enables motion-informed data re-acquisition, further promoting data consistency across scans and sites.

\subsection{Modality applicability and limitations}
Harmonized data acquisition and reconstruction are, in principle, applicable to nearly all MRI modalities, as they are inherently prospective approaches. Existing efforts have primarily focused on qMRI, dMRI, and fMRI, particularly those involving advanced pulse sequences \citep{RNMIMOSA,RNEPI,RNPTXPULSEQ,RNPULSEQX}. Vendor-neutral acquisition provides a unique avenue for disentangling the sources of variability in quantitative and functional MRI, enabling systematic investigation of whether observed differences arise from physiological factors or technical confounds. In addition, open-source frameworks facilitate rapid validation and dissemination of new methods, improve methodological transparency, and thereby promote reproducible cross-site studies. Taking Pulseq as an example, the community-driven Harmonized MRI\footnote{https://harmonizedmri.github.io/projects/} initiative has already collected dozens of open-source projects, and recent studies increasingly adopt the practice of publicly releasing their acquisition sequences and reconstruction pipelines alongside publication \citep{RNMIMOSA,RNEPI}.

Despite these advantages, the limitations of harmonized data acquisition and reconstruction should be critically examined. At a conceptual level, acquisition-level harmonization is inherently prospective and thus not applicable to most existing large-scale, retrospective MRI datasets, which significantly limits its real-world utility compared with post hoc strategies. In addition, current vendor-neutral pulse sequences often suffer from limited parameter flexibility and interactivity, making it difficult to adapt acquisition settings to subject-specific conditions or scanner constraints. This rigidity contrasts with vendor-native sequences that are typically highly optimized and dynamically adjustable. Moreover, the development and deployment of harmonized reconstruction frameworks remain heavily dependent on vendor cooperation, including access to low-level system interfaces and reconstruction pipelines. These barriers are not only technical but also institutional and regulatory, as clinical adoption requires extensive validation and approval. Importantly, large-scale multi-center validation studies are still lacking, with most existing works limited to novel or prototype pulse sequences or small cohorts, leaving scalability and generalizability insufficiently established. Overall, while acquisition-level harmonization is conceptually appealing, more work is needed before it can be effectively adopted in multi-site studies.

\section{Image-level Retrospective Harmonization}\label{sec:img}
\subsection{Traditional image-level approaches}\label{sec:imgt}
Early traditional image-level harmonization methods primarily rely on various intensity normalization techniques \citep{RN50, RN51}. Although some of these methods are not explicitly designed to remove site effects, they are commonly used as preprocessing steps or baseline approaches due to their simplicity and low computational cost. Such methods typically apply global transformations to the entire image (e.g., z-score normalization), adjust image intensity statistics (e.g., histogram matching), or utilize reference intensities from specific tissue types to align global or local intensity distributions, thereby improving comparability across scans acquired from different sites.

A widely used class of methods is based on histogram matching \citep{RN50,RN47}, which aims to align the intensity histogram or cumulative distribution function (CDF) of a source image with that of a target image or a predefined reference distribution. Although these methods are conceptually simple and computationally efficient, they are often sensitive to outliers (e.g., hyperintense lesions) and may fail to preserve biologically meaningful variations at the individual level. Another class of methods relies on reference-based normalization, such as White Stripe proposed by Shinohara et al. \citep{RN51}, which rescales the image intensities using a reference region composed of normal-appearing white matter (NAWM). Building upon this approach, RAVEL (Removal of Artificial Voxel Effect by Linear regression) \citep{RN46} further addresses residual non-biological variability that may persist after White Stripe normalization, which will be introduced in \Cref{sec:feas}. However, White Stripe relies on the assumption that NAWM serves as stable reference, which may not hold in populations with white matter pathology.

For diffusion MRI data, diffusion-weighted (DW) signals are often affected by differences in b-values, the number of diffusion gradient directions, angular sampling density, and global signal scaling induced by hardware or reconstruction/phase correction differences \citep{RN43,RN99}. These discrepancies cannot be adequately modeled as a simple intensity offset. A representative approach for dMRI is RISH (Rotationally Invariant Spherical Harmonics) \citep{RN48,RN45}, which decomposes voxel-wise DW signals into spherical harmonics for harmonization. In practice, RISH extracts spatially varying voxel-wise scaling factors by aligning RISH features across matched control groups and applies these factors to individual DW data. Due to its rotational invariance, RISH are robust to differences in gradient orientations and can serve as a preprocessing step compatible with a wide range of downstream analysis pipelines.  De Luca et al. \citep{RN49} further extended the RISH framework by introducing a covariate-driven general linear model (RISH-GLM), allowing multivariate modeling of site effects and cross-site harmonization without the need for matched reference data. Another representative approach is the Method of Moments (MoM) proposed by Huynh et al. \citep{RNMOM}. MoM derives voxel-wise scaling parameters by matching the spherical mean and variance of DW signals across sites and applies them to achieve signal-level harmonization. Compared with RISH, this method can be flexibly applied to datasets acquired with different numbers of gradient directions.

\subsection{Learning-based image-level approaches}\label{sec:imgl}
Learning-based image-level harmonization methods are predominantly driven by deep convolutional neural networks and can be broadly categorized into four groups: adversarial learning and style transfer, biological-site disentanglement, multi-contrast prior learning, and source-free distribution modeling. A general trend across these approaches is the shift from fixed, site-specific harmonization toward adaptive multi-site solutions, and from methods requiring simultaneous access to data from multiple sites to those leveraging only single-site data. In the context of deep learning, although MRI harmonization shares methodological similarities with domain adaptation and domain generalization \citep{RN151}, its purpose is not limited to improving model performance on a specific downstream task. As a result, its design principles, evaluation strategies, and intended outcomes are also fundamentally different.

\subsubsection{Adversarial learning and style transfer}
\begin{figure*}[t]
 \centering
 \includegraphics[width=1\textwidth]{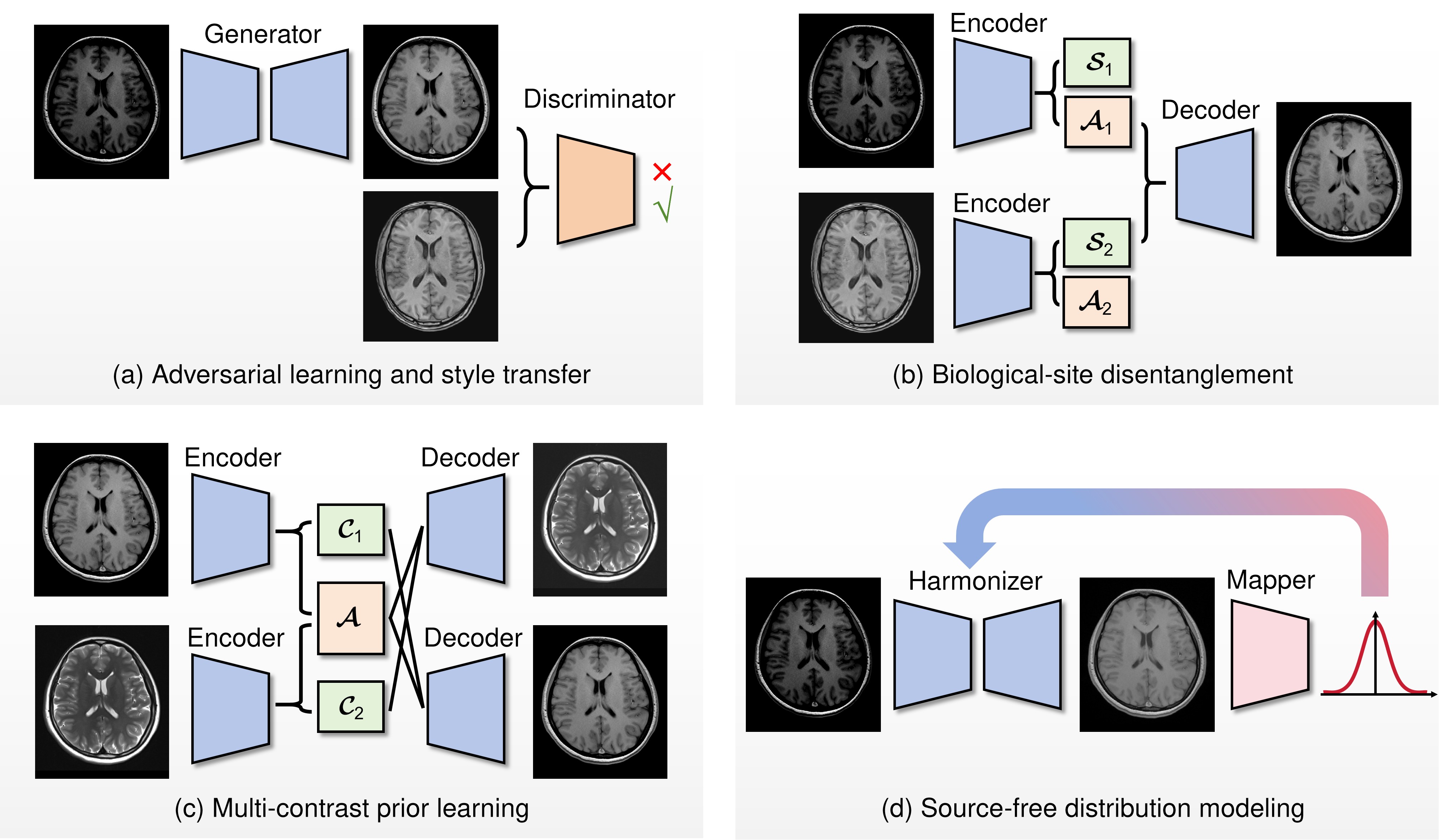}
 \caption{Four representative categories of image-level deep learning-based harmonization methods: (a) adversarial learning and style transfer, (b) biological-site disentanglement, (c) multi-contrast prior learning and (d) source-free distribution modeling. A: anatomy; S: style; C: contrast.}
 \label{fig:fig3}
\end{figure*}

For learning-based approaches, image-level harmonization is closely related to image style transfer, with generative adversarial networks (GANs), particularly CycleGAN, being among the earliest methods for unpaired harmonization (\Cref{fig:fig3}a) \citep{RN54,RN56,RN53,RN55,RN58}. These approaches treat data from different sites as source and target domains and use cycle-consistency constraints to transfer appearance while preserving anatomy. However, CycleGAN requires separate models for each site pair and cannot exploit shared information across multiple sites. To address this limitation, many-to-one GAN-based strategies have been proposed. For example, IGUANe (Image Generation with Unified Adversarial Networks) introduces a universal generator to map images from multiple sites to a reference domain, trained with multiple site-specific discriminators and backward generators to enforce cycle consistency, albeit with substantial computational overhead \citep{RN57}. More recent frameworks, such as StyleGAN and StarGAN, incorporate explicit site or style encoding to improve scalability and reduce model complexity \citep{RN59,RN64,RN52,RN61}.

As early Image-level learning-based harmonization approaches, the risks associated with GAN-based methods warrant careful consideration, particularly in medical MRI applications. Due to the lack of explicit biological fidelity constraints and the reliance on distribution-level alignment, GAN-based methods are prone to hallucination effects, where subtle biologically meaningful variations may be introduced or suppressed. Such changes can potentially affect diagnosis and downstream analyses, yet remain difficult to detect with existing evaluation metrics. In addition, the inherent instability of GAN training complicates reproducibility and hyperparameter tuning in multi-site settings. Furthermore, given the limited size and high inter-sample similarity of MRI datasets, GAN-based models are susceptible to mode collapse, which may lead to the loss of subject-specific biological information during harmonization. Although methods such as CycleGAN and StyleGAN partially alleviate these limitations, they do not fully eliminate them. These considerations highlight the need for careful validation of GAN-based harmonization methods.

\subsubsection{Biological-site disentanglement}
Image style transfer provides a straightforward solution for harmonization but lacks explicit separation between biological and site-related factors. To address this limitation, early methods based on variational autoencoders (VAEs) with adversarial learning, enable more structured modeling by encoding images into latent representations, allowing explicit disentanglement of biological information and site-related variation. In the context of structural MRI, these factors are often instantiated as anatomical content and image style, respectively (\Cref{fig:fig3}b). For example, MURD introduces separate encoders for anatomical and style information and achieves harmonization by recombining source anatomy with target style. The traveling human phantom dataset demonstrates that the MURD method achieves comprehensive improvements over GAN-based approaches in both quantitative imaging metrics and downstream segmentation tasks, with gains exceeding 20\% on average \citep{RN68}. ImUnity simplifies this framework using contrastive learning to encourage anatomical consistency. Compared with CycleGAN, it improves SSIM from 0.87 to 0.95 on traveling-subject data after harmonization, indicating enhanced preservation of anatomical structures \citep{RN73}. However, as these approaches fundamentally rely on adversarial learning, they remain susceptible to issues such as instability and mode collapse.

Recent approaches have turned to diffusion models as a more stable generative framework for harmonization. They formulate image translation as a progressive noise perturbation and denoising process, enabling more stable training and higher-fidelity generation \citep{RNdiff1,RNdiff2}. For example, Lan et al. formulated harmonization as a controllable domain adaptation problem, where a domain-invariant anatomical condition is learned and domain embeddings steer the denoising trajectories of a single diffusion model for flexible multi-site harmonization \citep{RNfea03}. HCLD (Harmonization framework through Conditional Latent Diffusion) further improves efficiency by performing diffusion in a compressed latent space \citep{RNfea04}. Advances from image translation and synthesis may  offer additional insights. For instance, diffusion bridges reformulate the generation process from noise-to-image into an image-to-image translation task, which is more closely aligned with the nature of harmonization \citep{RNfea02}. In addition, flow matching directly approximates the transport path between noise and data distributions, significantly accelerating conventional diffusion models \citep{RNfea05}.

Unlike explicit disentanglement strategies, implicit approaches do not impose hard constraints to separate content and site-specific style. Instead, they allow the model to allocate part of its representational capacity to capture site-specific attributes while maintaining a shared subspace for site-invariant anatomical information. A representative example is pFLSynth, which adopts label-guided conditioning to enable controllable harmonization, avoiding aggressive suppression of site-related variations and thereby reducing the risk of anatomical degradation \citep{RNfea01}. By preserving a shared latent subspace, these models can effectively leverage previously learned knowledge during fine-tuning on unseen sites, leading to improved generalization.

\subsubsection{Multi-contrast prior learning}
Clinical acquisitions and a number of large-scale public MRI datasets often include multiple contrasts per subject to capture complementary tissue properties. These contrasts are commonly assumed to share consistent or similar anatomical structures, thereby providing naturally paired anatomy-contrast information for disentanglement (\Cref{fig:fig3}c). In practice, even a subset of available contrasts can be effectively leveraged to exploit shared anatomical information. Beyond structural consistency, multi-contrast MRI is intrinsically linked through shared tissue properties (e.g., T1 and T2 relaxation times), offering a unified representation of underlying tissue characteristics across contrasts. Together, these properties form a strong foundation for multi-contrast prior learning in harmonization.

From the perspective of structural consistency, many approaches aim to explicitly disentangle latent representations using co-registered modalities. For example, Dewey et al. leveraged co-registered T1- and T2-weighted images from the same subject to disentangle latent representations, enabling harmonization by combining images from a new site with a reference contrast \citep{RN157}. Building upon this, CALAMITI introduces cross-contrast synthesis and adversarial learning to enhance disentanglement and enforce globally consistent anatomical representations across sites \citep{RN66}. HACA3 further challenges the assumption of identical anatomy in MR disentanglement by leveraging contrastive learning to preserve inherent anatomical differences, enabling flexible contrast combinations and improved robustness to incomplete or heterogeneous data \citep{RN69}. In parallel, physics-driven approaches leverage shared tissue properties by mapping multi-contrast images to modality- and protocol-invariant quantitative parameters \citep{RN70,RN71}. These methods combine physical forward models to enable self-supervised learning, with extensions such as PhyCHarm further integrating scanner-specific acquisition parameters to synthesize parametric maps for harmonization, although paired supervision is still required in the final stage \citep{RN156}.

Despite these advances, the applicability of multi-contrast prior learning remains relatively limited. Multi-contrast data from the same subject is not always available in multi-site datasets and, if available, may not perfectly align due to potential subject motion between scans. This issue is particularly pronounced in certain populations and anatomical regions (e.g., fetal and neonatal imaging, as well as cardiac and abdominal imaging). As a result, such methods may exhibit reduced flexibility and transferability, and be less robust to missing or corrupted contrasts.

\subsubsection{Source-free distribution modeling}
To eliminate the dependency on multi-contrast and multi-site data while ensuring generalizability to unseen domains, normalizing flows have been introduced to directly model the source distribution \citep{RN67,RN74}. Unlike GAN-based approaches that often suffer from mode collapse, flows provide a more principled, likelihood-based objective for density estimation. This enables source-free harmonization without the need for traveling subjects, multi-site data, or task-specific supervision. By mapping a complex probability distribution to a simple latent space through a series of invertible and differentiable transformations, flows ensure a bijective mapping between domains. This inherent invertibility offers advantages for medical imaging, as it may help preserve fine-grained anatomical details and reduce the risk of hallucinations or information loss commonly observed in GAN-driven synthesis.

Jeong et al. and Beizaee et al. independently introduced normalizing flows into image harmonization, proposing the BlindHarmony \citep{RN67} and Harmonizing Flows \citep{RN74} frameworks, respectively. Taking Harmonizing Flows as an example, the method follows a three-step strategy: source domain modeling, harmonizer pre-training, and test-time adaptation. A normalizing flow model, composed of stacked affine coupling layers,  is first trained to capture the source distribution and map it to a standard Gaussian via invertible transformations. A lightweight U-Net is then pre-trained to reconstruct source images from augmented inputs, learning to compensate for appearance variations. During inference, the harmonizer is fine-tuned on target data under the supervision of the frozen flow model, aligning outputs with the source distribution (\Cref{fig:fig3}d). This framework enables unsupervised, source-free, and task-agnostic harmonization, and demonstrates strong generalization to unseen domains across tasks such as brain MRI segmentation and neonatal age estimation.

\subsection{Modality applicability and limitations}
Image-based harmonization methods have been predominantly developed for conventional structural MRI contrasts, such as T1, T2, PD (proton density), and FLAIR (fluid-attenuated inversion recovery) images. Although a limited number of studies have claimed potential generalizability to other MRI modalities, substantive evidence supporting such claims remains scarce. This lack of validation underscores the limited generalizability and cross-modality transferability of image-based harmonization approaches. In the context of dMRI, harmonization efforts are dominated by RISH-based methods and their extensions, whereas learning-based approaches largely rely on traveling-subject data for supervision (\Cref{sec:tsl}). In contrast, to date, no dedicated image-based harmonization methods have been established for fMRI or qMRI.

Beyond their limited generalizability, image-based harmonization methods are also prone to over-correction and the removal of biologically meaningful variability. This risk primarily arises from the lack of explicit biological or covariate constraints and the failure to account for the confounding between site effects and biological variables. While learning-based approaches have substantially advanced harmonization performance in quantitative metrics, these gains may come at the cost of reduced interpretability, increased dependence on large-scale training data and computational resources. Such methods may also introduce hallucination effects that are inherently challenging to detect and avoid. This limitation is exacerbated by the fact that existing validation strategies largely rely on qualitative visual comparisons of samples outside the training data \citep{RN74}. Furthermore, image-based methods are sensitive to preprocessing pipelines and registration accuracy, and their harmonization outcomes are often strongly dependent on the choice of reference site.

\section{Feature-level Retrospective Harmonization}\label{sec:fea}
\subsection{Statistical approaches}\label{sec:feas}
Feature-level methods based on statistical modeling typically assume that extracted features (e.g., brain volumes, cortical thickness, dMRI metrics) can be decomposed into biological effects, site or batch effects, and random noise. By fitting a statistical model (most commonly a linear model), the site effect can be estimated and subsequently removed or adjusted, yielding harmonized data. A comprehensive review of such methods is available in Hu et al.\citep{RN39} Here, we briefly introduce several representative approaches, with a particular emphasis on ComBat \citep{RN99,RN77}, which serves as a foundation for subsequent learning-based methods.

The ComBat model was originally developed for batch effect correction in gene expression data \citep{RN77}, and was first introduced to dMRI data \citep{RN99} In this model, the observed feature $y_{ijv}$ at voxel $v$ for subject $j$ from site $i$ is modeled as a linear combination of multiple factors. These factors include the global mean $\alpha_{v}$, biological covariates (e.g., age and sex), and site effects (additive and multiplicative effects, $\gamma_{iv}$ and $\delta_{iv}$). The full model can therefore be expressed as:
\begin{equation}
y_{ijv} = a_v + \mathbf{X}_{ij} \beta_v + \gamma_{iv} + \delta_{iv} \varepsilon_{ijv}
\label{eq:eq1}
\end{equation}
where $\mathbf{X}$ is the design matrix for the covariates, $\beta$ represents the corresponding regression coefficients, and $\varepsilon$ is the error term, assumed to have zero mean and variance $\sigma^{2}$. After estimating the coefficients using the empirical Bayes method, the ComBat-harmonized value can be expressed as:
\begin{equation}
y_{ijv}^{\mathrm{ComBat}} = 
\frac{y_{ijv} - \hat{a}_v - \mathbf{X}_{ij} \hat{\beta}_v - \hat{\gamma}_{iv}}{\hat{\delta}_{iv}} 
+ \hat{a}_v + \mathbf{X}_{ij} \hat{\beta}_v
\label{eq:combat}
\end{equation}
Based on this, ComBat has been shown to be effective across a variety of imaging features, including not only dMRI-derived metrics but also cortical thickness, functional connectivity, quantitative tissue parameters, spectroscopy and radiomics \citep{RN86,RN95,RN91,RN88,RN97,RN92,RN93,RNqmri}. It is worth noting that, although ComBat can also be applied after normalizing images to a standard space (image-level harmonization), this is often a suboptimal choice and is therefore typically used only as a comparative baseline \citep{RN74,RNHACA3}.

Based on the standard ComBat, numerous extensions have been proposed, including the use of alternative parameter estimation strategies and applications to more complex study designs \citep{RN82,RN101,RN78,RN81,RN75,RN96,RN98,RN131}. For example, ComBat-GAM introduces generalized additive models (GAMs) to model nonlinear covariate effects on feature means \citep{RN82}. ComBatLS further generalizes this approach using GAMLSS (generalized additive models for location, scale, and shape) to account for covariate-dependent variability in both mean and variance \citep{RNLS}. However, one major limitation of ComBat-based methods is their reliance on sufficient within-scanner samples to estimate site-specific effects. This constraint reduces its generalizability to unseen data \citep{RN5}.

Unlike ComBat, which explicitly models additive and multiplicative effects, some other strategies model biological factors or site effects using basis representation or latent factors \citep{RN46,RN87,RN94,RN85,RN120}. Taking CovBat \citep{RN94} as an example, it extends ComBat by explicitly accounting for site effects in covariance. CovBat performs principal component analysis on the ComBat-adjusted residuals and conducts harmonization in the principal component space, yielding CovBat residuals, $\hat{\varepsilon}_{ijv}^{\mathrm{CovBat}}$. Consequently, Eq. (6) is reformulated as:
\begin{equation}
y_{ijv}^{\mathrm{CovBat}} = \hat{\varepsilon}_{ijv}^{\mathrm{CovBat}}
+ \hat{a}_v + \mathbf{X}_{ij} \hat{\beta}_v
\label{eq:combat}
\end{equation} 
Similarly, Zhang et al. proposed FELIEF (REmoval of Latent Inter-scanner Effects through Factorization) \citep{RN120}, which addresses more complex site effects by explicitly modeling scanner-related latent multivariate structures. The method applies matrix factorization to standardized residuals and imposes low-rank constraints on the latent components via nuclear norm regularization, thereby effectively identifying and removing scanner-related technical variability. Another representative example is RAVEL \citep{RN46}, which selects a control voxel that is highly sensitive to variations in reconstruction algorithms, acquisition protocols, and scanner configurations, typically from the cerebrospinal fluid, to serve as a proxy for non-biological effects. Then, RAVEL performs singular value decomposition on the control voxels to extract latent factors representing technical variation, and then applies linear regression across all voxels to estimate and remove these effects. Finally, the resulting residuals are treated as the RAVEL-corrected intensities. 

\subsection{Learning-based feature-level approaches}\label{sec:feal}
Learning-based feature-level approaches are typically extensions of statistical harmonization strategies, particularly those derived from ComBat. A representative example is Neuroharmony \citep{RN1,RN4}, which is based on the assumption that the intrinsic image characteristics of a single image can aid in data harmonization. This approach addresses the limitation of traditional ComBat, which cannot generalize to unseen sites. Specifically, Neuroharmony first applies ComBat to existing multi-site data to obtain corrected features for each image. Then, using the MRIQC tool, a range of image quality metrics (IQMs) are extracted from each image, including SNR, contrast, blurriness, motion artifacts, and background uniformity. Based on these metrics, Neuroharmony employs a random forest model to learn the mapping between the 64 IQMs and the ComBat-derived corrected features. Once trained, the model no longer relies on population-level statistical features but instead performs harmonization using only the image’s IQMs and biological covariates.

Another line of learning-based feature-level approaches leverages conditional variational autoencoder (cVAE) \citep{RN7} to address nonlinear site-related variations and support multivariate modeling. In the cVAE framework, an encoder first processes feature vectors to generate latent representations. These representations are then concatenated with site information (i.e., a one-hot vector) or biological covariates and fed into a decoder to reconstruct the original feature vectors. To accommodate 1D input, both the encoder and decoder are typically implemented as fully-connected neural networks. To encourage site-invariant latent representations, mutual information between the latent features and the site encodings is minimized during training. Then, in the harmonization phase, modifying the input site encoding allows for flexible translation of input features to any target site. 

Several extensions of the cVAE framework have been developed to enhance harmonization performance. For instance, the goal-specific cVAE (gcVAE) \citep{RN2} proposed by An et al. incorporates a pretrained classifier into the standard cVAE architecture. This allows the original cVAE to implicitly preserve biologically meaningful representations by leveraging supervision from downstream classification tasks during training. Another variant, DeepComBat by Hu et al. \citep{RN3}, integrates cVAE with the classical ComBat method. It first applies ComBat to the latent mean vectors produced by the cVAE encoder, followed by decoding the harmonized latent representations to reconstruct the original features. A second ComBat step is then applied to the residuals (i.e., the difference between the reconstructed and original features) to remove residual site effects, which are subsequently added back to produce the final harmonized output.

\begin{figure*}[t]
 \centering
 \includegraphics[width=0.8\textwidth]{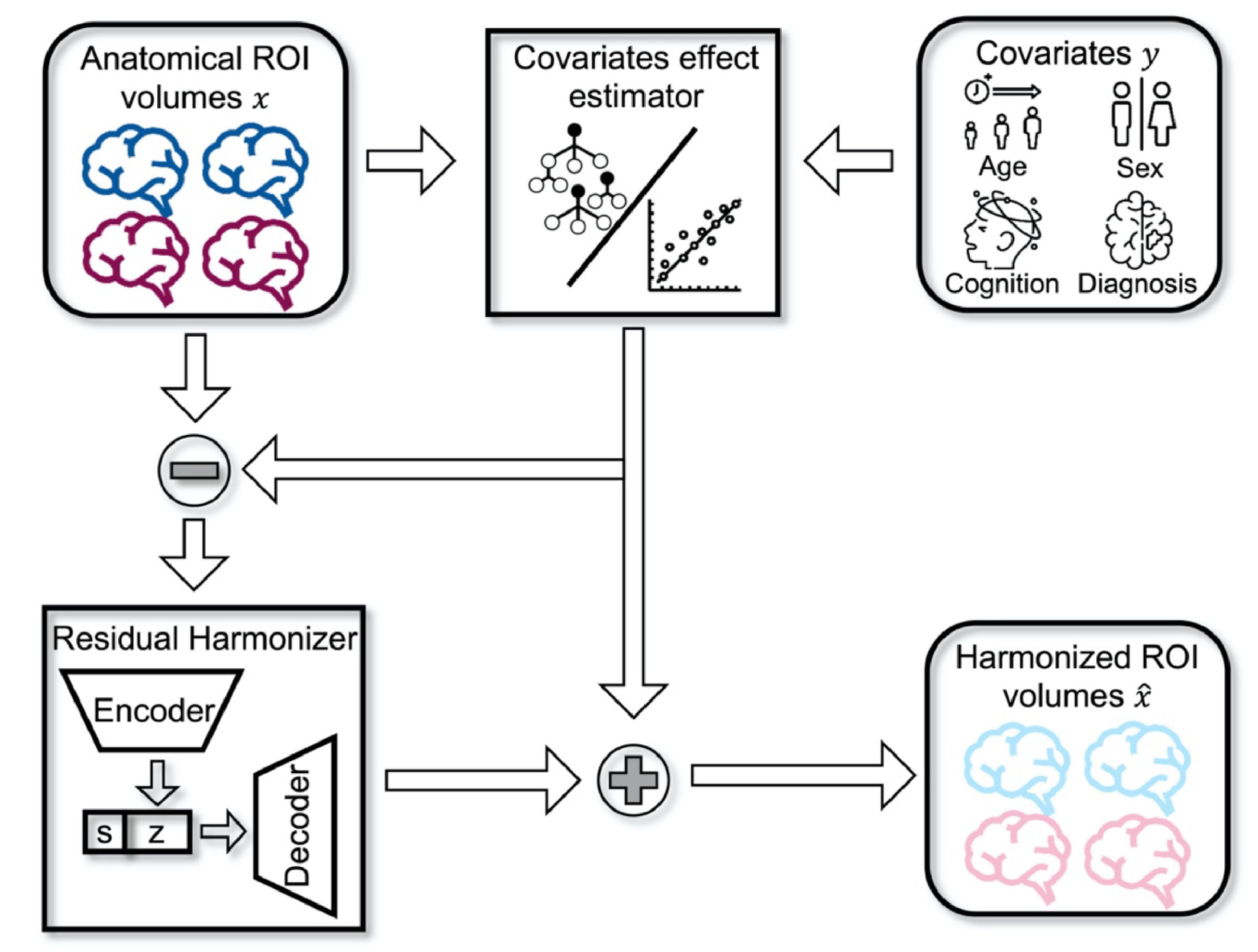}
 \caption{Schematic illustration of the feature-level deep learning harmonization method DeepResBat \citep{RN5}. The covariates effect of the original features were first removed by subtraction and used as the input to a VAE. Then, the VAE output was added back to the removed covariate components to obtain harmonized features. ROI: region of interest.}
 \label{fig:fig4}
\end{figure*}

Distribution differences in covariates (e.g., age and sex) are common and often unavoidable in multi-site datasets. As theoretically demonstrated by Tachet et al. \citep{RN155}, directly applying cVAE under such conditions may lead to covariate-driven variations being incorrectly attributed to site effects. To address this issue, DeepResBat \citep{RN5} introduces a two-stage strategy. Specifically, it first estimates covariate influences using nonlinear regression models. The covariate residuals, obtained by subtracting the estimated covariate contributions from the original features, are then used as input to a cVAE model to isolate and remove site-specific effects, yielding harmonized residuals. The final harmonized features can be reconstructed by reintroducing the covariate effects into the harmonized residuals (\Cref{fig:fig4}). By targeting residuals rather than raw features for deep learning harmonization, DeepResBat explicitly preserves biological variability while effectively reduces the risk of spurious associations.

\subsection{Modality applicability and limitations}
Feature-level methods operate on MRI-derived measurements or features rather than raw image intensities, and therefore are applicable to nearly all MRI modalities and are particularly well suited for large-scale, existing multi-center MRI datasets. Their flexibility, relatively strong generalizability, interpretability, and modest computational requirements have contributed to their continued status as the most widely adopted harmonization paradigm to date. As discussed above, ComBat and its extensions have been successfully applied to a broad range of feature types, including cortical thickness (structural MRI) \citep{RN95}, fractional anisotropy (dMRI) \citep{RN99}, functional connectivity (fMRI) \citep{RN86}, and even metabolite concentrations (MR spectroscopy) \citep{RN92}. Owing to their algorithmic simplicity and practical feasibility, feature-level harmonization methods have been supported by well-established implementations, including ComBatHarmonization\footnote{https://github.com/Jfortin1/ComBatHarmonization/}, neuroHarmonize\footnote{https://github.com/rpomponio/neuroHarmonize}, and DPABI harmonization\footnote{https://rfmri.org/content/dpabi-harmonization-toolbox-harmonizing-multi-site-brain-imaging-big-data-era}, thereby facilitating their broad adoption in large-scale and translational studies.

Unlike image-level harmonization, feature-level harmonization is typically task-driven, aiming to improve the stability and cross-site generalization of downstream models by reducing site-induced shifts in feature distributions. However, because it directly modifies the measurements used for statistical analysis and predictive modeling, its impact extends beyond removing site effects to influencing the preservation of biologically meaningful variation, thereby increasing the risk of overcorrection in the presence of confounding. In addition, feature-level methods rely on a set of statistical assumptions, including linear or parametric formulations of site effects and the assumption that site-effect parameters across features are drawn from shared prior distributions \citep{RN99,RN94,RN5}. Such assumptions may fail to capture spatially varying site effects, which in MRI are often introduced by field inhomogeneities and gradient nonlinearity. Furthermore, feature-level approaches are highly dependent on feature definitions and upstream processing pipelines: systematic biases introduced during preprocessing cannot be corrected by subsequent harmonization, leading to limited comparability of derived features across studies employing different processing strategies. From an outcome perspective, feature-level methods do not produce harmonized images, which constrains their task-agnostic scalability and broader applicability.

\section{Traveling Subject}\label{sec:ts}
Harmonization models based on non-traveling subject datasets, whether traditional statistical approaches or learning-based methods, can effectively eliminate site effects. However, it remains unclear whether such models may also overcorrect biological variability. In this context, traveling subject-based approaches offer a baseline for rigorously disentangling biological and non-biological sources of variability. Moreover, publicly available traveling subject datasets not only enable investigation into how site effects influence multi-site statistical analyses, but also serve as valuable benchmarks for validating newly proposed harmonization methods.

\subsection{Statistical approaches}
Building on image-based histogram matching methods, Wrobel et al. proposed Multisite Image Harmonization by Cumulative Distribution Function Alignment (MICA) \citep{RN90}, which performs image harmonization based on the alignment of CDFs. The method first applies preprocessing steps such as N4 bias field correction and skull stripping to the images, and then computes their CDFs. For each traveling subject, MICA selects one image as the template and uses its CDF as the alignment target. It then estimates a nonlinear, monotonically increasing warping function by densely sampling paired points between the source and template images and applying linear interpolation. This warping function is used to align the CDF of the source image to that of the template.

Based on traveling subject (TS) data, ComBat can also be extended to the TS-ComBat method \citep{RN83}. In this approach, the covariate term in the standard ComBat model is replaced with individual effects estimated from traveling subjects, while the site effects are still estimated and removed using an empirical Bayes method. To further account for repeated measurements across time points, Beer et al. \citep{RN89} applied Longitudinal ComBat in the context of traveling subject studies. The model is expressed as follows:
\begin{equation}
y_{ijv}(t) = a_v + \mathbf{X}_{ij}(t) \beta_v + \eta_{jv} + \gamma_{iv} + \delta_{iv} \varepsilon_{ijv}(t)
\label{eq:longitudinal}
\end{equation}
where both $y_{ijv}(t)$ and $\varepsilon_{ijv}(t)$ introduced time-varying dependent variables, $\eta_{jv}$ represents subject-specific random intercept. In addition, several of the previously discussed image-level and feature-level harmonization methods have been further validated and extended on traveling subject datasets, demonstrating their adaptability across sites \citep{RN83,RN76,RN80}.

\subsection{Learning-based approaches} \label{sec:tsl}
\begin{figure*}[t]
 \centering
 \includegraphics[width=1\textwidth]{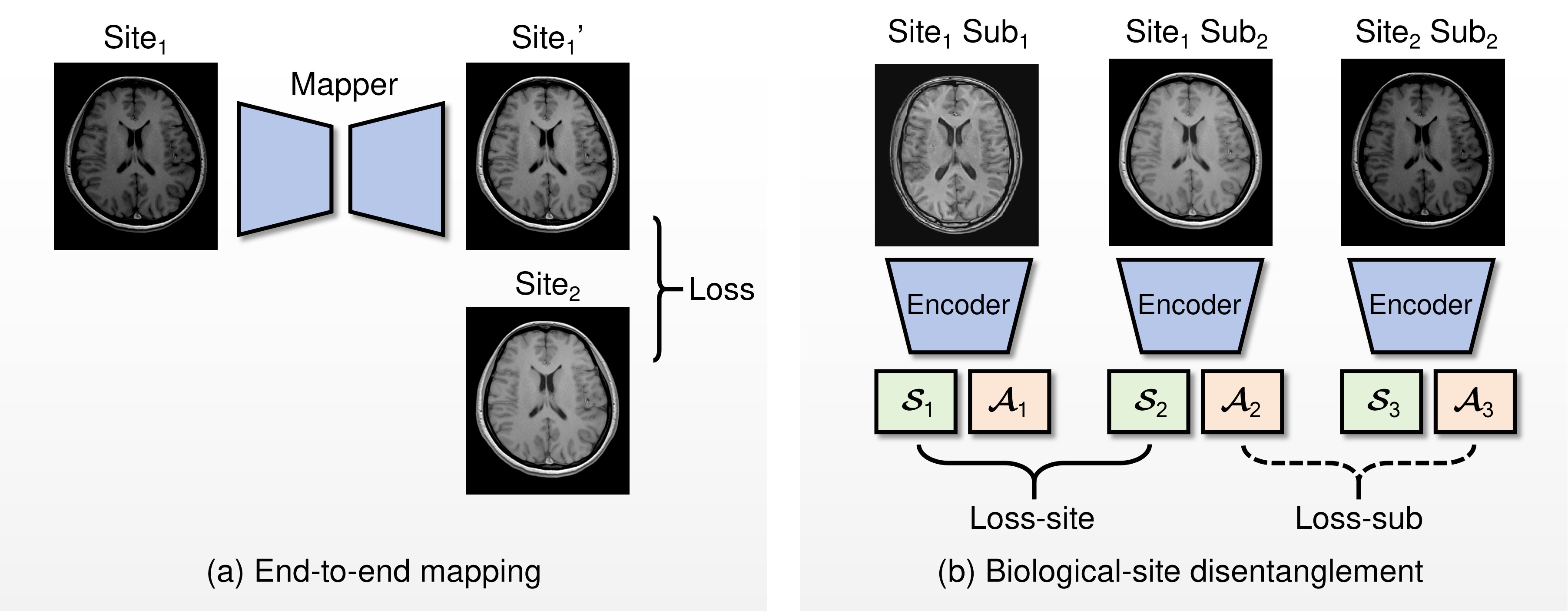}
 \caption{Two representative deep learning-based harmonization strategies using traveling subject data: (a) end-to-end mapping and (b) biological-site disentanglement methods. The availability of paired training data provides additional supervision related to site or subject identity, which enhances the learning of site-invariant representations. A: anatomy; S: style.}
 \label{fig:fig5}
\end{figure*}

Traveling-subject data naturally provide paired training samples for learning-based methods, offering a more direct solution compared to unpaired approaches. Methodologically, these techniques can be categorized into two main types: end-to-end mapping and biological-site disentanglement strategies similar to those used in unpaired settings (\Cref{fig:fig5}).

\subsubsection{End-to-end mapping}
Based on paired training data, Dewey et al. proposed DeepHarmony \citep{RN36}, a U-Net-based harmonization framework. In their study, 12 subjects were scanned on two different scanners using protocols (e.g., T1, T2, PD, and FLAIR) with varying parameters. U-Net was then directly trained to learn an image-to-image mapping between paired images acquired from the two sites. DeepHarmony was shown to substantially reduce inter-protocol volumetric discrepancies in longitudinal MRI datasets of patients with multiple sclerosis.

Unlike methods that perform direct mapping in the image domain, Tong et al. \citep{RN31} proposed a harmonization approach that maps source DW images to target diffusion kurtosis imaging (DKI) parameters. A 3D hierarchical convolutional neural network was trained using co-registered labels estimated through an iteratively reweighted linear least squares method. This approach resulted in a 50-60\% reduction in inter-scanner variation of DKI parameters within white matter. Similarly, Tax et al. \citep{RN35} and Ning et al. \citep{RN6} summarized several learning-based harmonization methods from the Multi-Shell Diffusion MRI Harmonization Challenge (MUSHAC). These methods harmonize dMRI data in the spherical harmonics domain. All included methods significantly reduced variability across multi-scanner dMRI acquisitions, although challenges remain in accurately capturing localized features.

To improve generalizability to unseen sites, Xu et al. proposed Site Mix (SiMix) \citep{RN33}, which combines mixed-site training with test-time perturbation. Instead of harmonizing to a single existing site, SiMix constructs a virtual target site by linearly combining images from multiple known sites during training. At inference, the test image is mixed with its initially harmonized output to generate multiple perturbed inputs, whose predictions are averaged to produce the final harmonized result, following an ensemble strategy.

\subsubsection{Biological-site disentanglement}
Compared to non-traveling-subject methods, traveling-subject-based biological-site disentanglement offers the distinct advantage of efficiently utilizing shared anatomical structures across different sites for strong supervision, thereby enabling more accurate interpretation and quantification of site effects (\Cref{fig:fig5}b).

A representative method is Multi-scanner Image harmonization via Structure Preserving Embedding Learning (MISPEL), proposed by Torbati et al. \citep{RN32} The framework consists of scanner-specific encoders and decoders and follows a two-stage training strategy.  This approach demonstrates that paired multi-site data can provide strong supervision, enabling the model to maintain high anatomical fidelity during harmonization. Notably, ESPA, proposed by Torbati et al. \citep{RN38} and built on the MISPEL framework, relaxes the requirement for traveling-subject paired data by employing augmentation strategies on single-site images. Additionally, Tian et al. \citep{RN34} proposed a bidirectional framework called deep learning-based representation disentanglement (DeRed). This framework consists of four encoders to disentangle anatomical and site-specific representations from paired different sites, and two decoders to bidirectionally synthesize harmonized images. A key advantage of this model is its flexible multi-site harmonization capability, where new unseen sites can be linked to the target site via intermediate domains without retraining the entire model.

\subsection{Available traveling subject datasets}
One of the major challenges in image harmonization is how to effectively evaluate its performance.  A direct and reliable approach is to use traveling-subject datasets, which minimize the bias introduced by inter-site population sampling. However, acquiring such datasets is often costly and limited by the number of available participants. Therefore, leveraging existing publicly available traveling-subject datasets is often helpful. \Cref{tab:tab2} summarizes the currently available public datasets, covering traveling subjects across different imaging modalities and age groups, and involving major scanner vendors or varying acquisition protocols \citep{RN35,RN118,RN154,RN153}.

\begin{table*}[!ht]
    \captionsetup{font=normalsize} 
    \centering
    \caption{Available traveling subject dataset}
    \label{tab:tab2}
    \small
    \begin{tabular}{L{0.17\textwidth} L{0.11\textwidth} L{0.11\textwidth} L{0.15\textwidth} L{0.33\textwidth}}  
        \toprule
        \textbf{Dataset name} & \textbf{Number of subjects (age)} & \textbf{Number of Scanners/sites}  & \textbf{MRI modalities} & \textbf{Data Repository}\\ 
        \midrule
        \makecell[l]{ON-Harmony \\\citep{RN118}}&{20 (18-55y)}&{6/5}&{T1w, T2w, SWI, dMRI, rs-fMRI}&{https://openneuro.org/datasets/ds004712}\\
        \midrule
        \makecell[l]{SRPBS \\\citep{RN153}}&{9 (24–32y)}&{12/8}&{T1w, rs-fMRI, fieldmap}&{https://bicr-resource.atr.jp/srpbsts}\\
        \midrule
        \makecell[l]{SDSU-TS \\\citep{hau2025traveling}}&{9 (22-55 y)}&{2/2}&{T1w, T2w, dMRI}&{https://openneuro.org/datasets/ds005664}\\
        \midrule
        \makecell[l]{HAMLET}&{5 (N/R)}&{4/3}&{T1w, dMRI, rs-fMRI}&{https://www.nitrc.org/projects/hamlet}\\
        \midrule
        \makecell[l]{ZJU dMRI \\\citep{RN154}}&{3 (23-26 y)}&{10/10}&{T1w, dMRI}&{https://doi.org/10.6084/m9.figshare.8851955.v6}\\
        \midrule
        \makecell[l]{SPINS Human \\Phantoms \\\citep{HAWCO2018134}}&{4 (N/R)}&{6/3}&{T1w, dMRI, rs-fMRI}&{https://openneuro.org/datasets/ds003011}\\
        \midrule
        \makecell[l]{MUSHAC \\\citep{RN35}}&{14 (21-41 y)}&{3/ (N/R)}&{dMRI}&{https://www.cardiff.ac.uk/cardiff-university-brain-research-imaging-centre/research/projects/cross-scanner-and-cross-protocol-diffusion-MRI-data-harmonisation}\\
        \bottomrule
        \makecell[l]{$^{*}$N/R: Not reported}
    \end{tabular}
\end{table*}

\begin{figure*}[t]
 \centering
 \includegraphics[width=0.9\textwidth]{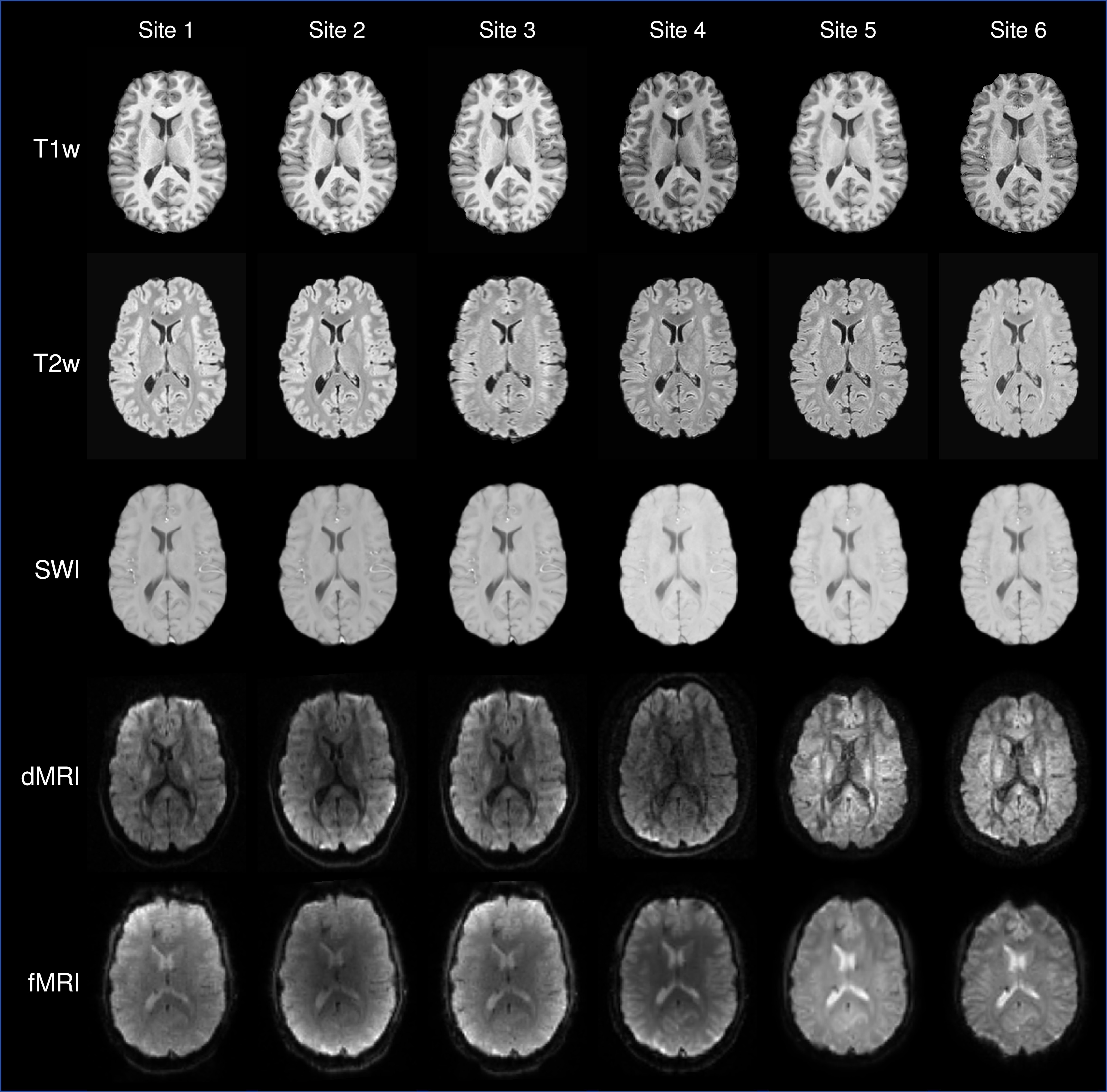}
 \caption{Representative examples of different modalities for a single participant data across all scanners from ON-Harmony dataset \citep{RN118}.}
 \label{fig:fig6}
\end{figure*}

Taking ON-Harmony \citep{RN118} as an example, 20 healthy volunteers were scanned using five imaging modalities across six scanners from different vendors and models. These modalities included structural imaging (T1-weighted, T2-weighted, and susceptibility-weighted imaging) as well as functional imaging (dMRI and resting-state fMRI). As shown in \Cref{fig:fig6}, a clear observation is that functional modalities exhibit substantially greater inter-scanner variability than structural ones. This discrepancy arises not only from differences in reconstruction and post-processing pipelines across scanners, but also from the fact that both dMRI and fMRI typically rely on fast echo-planar imaging sequences for data acquisition, which are more susceptible to imperfections such as field inhomogeneities.

\section{Evaluation Metrics}\label{sec:em}
The challenge of validating harmonization remains a central bottleneck in the field. The absence of a definitive ground truth, together with the limitations of existing evaluation metrics, often renders validation even more challenging than harmonization itself. As a result, objective method comparison becomes challenging, and clinical translation is hindered. In this section, we review three categories of validation strategies reported in the literature. A comparative synthesis of MRI harmonization method families and their corresponding validation paradigms is summarized in \Cref{tab:tab3}. It should be noted that none of these strategies alone can conclusively establish biological fidelity. Instead, they assess different aspects of harmonization effectiveness through complementary yet inherently limited evaluation perspectives.

\begin{table*}[!ht]
    \captionsetup{font=normalsize} 
    \centering
    \caption{Comparative synthesis of MRI harmonization method families}
    \label{tab:tab3}
    \small
    \begin{tabular}{M{0.15\textwidth} M{0.13\textwidth} M{0.2\textwidth} M{0.25\textwidth} M{0.15\textwidth}}  
        \toprule
        \textbf{Harmonization level} 
        & \textbf{Primary object of harmonization} 
        & \textbf{Advantages}  
        & \textbf{Weaknesses} 
        & \textbf{Typical validation}\\ 
        
        \midrule 
        \makecell[l]{Acquisition or \\reconstruction-level \\(\Cref{sec:acq})}
        & \cell{Signal formation and reconstruction}
        & \cell{%
            \tabitem Controls variability at the signal source\\
            \tabitem High interpretability}
        & \cell{%
            \tabitem Not applicable to retrospective datasets\\
            \tabitem Depends on vendor support\\
            \tabitem Challenging for large-scale deployment}
        & \cell{
            \tabitem Traveling subjects\\ 
            \tabitem Phantoms}\\
            
        \midrule 
        \makecell[l]{Image-level \\(traditional) \\(\Cref{sec:imgt})}
        & \cell{Image intensity or signal representation statistics}
        & \cell{%
            \tabitem Simple and efficient\\
            \tabitem No training required\\
            \tabitem Broad applicability}
        & \cell{%
            \tabitem Simple intensity transformations\\
            \tabitem Limited for complex site effects\\
            \tabitem Risk of removing biological variability}
        & \multirow{2}{=}{\cell{%
            \tabitem Traveling subjects\\
            \tabitem Visual assessment\\
            \tabitem Distributional statistics\\
            \tabitem Site discriminability tests\\
            \tabitem Downstream tasks performance}}\\

        \cmidrule(lr){1-4} 
        \makecell[l]{Image-level \\(learning-based) \\(\Cref{sec:imgl})}
        & \cell{Image appearance or contrast}
        & \cell{%
            \tabitem Strong visual harmonization\\
            \tabitem Models complex site effects}
        & \cell{%
            \tabitem Limited interpretability\\
            \tabitem Weak biological constraints\\
            \tabitem Risk of over-correction or hallucination\\
            \tabitem Data intensive}
        & {}\\
        
        \midrule 
        \makecell[l]{Feature-level \\(statistical) \\(\Cref{sec:feas})}
        & \cell{Image-derived features}
        & \cell{%
            \tabitem Explicit modeling of biological covariates\\
            \tabitem Statistically interpretable\\
            \tabitem Well established}
        & \cell{%
            \tabitem Relies on simple assumptions\\
            \tabitem Limited capacity for nonlinear or high-dimensional site effects\\
            \tabitem Sensitive to feature definition and sample imbalance\\
            \tabitem No harmonized images}
        & \multirow{2}{=}{\cell{%
            \tabitem Biological association analysis\\
            \tabitem Robustness tests under confounding\\
            \tabitem Site discriminability tests\\
            \tabitem Downstream tasks performance}}\\

        \cmidrule(lr){1-4} 
        \makecell[l]{Feature-level \\(learning-based) \\(\Cref{sec:feal})}
        & \cell{Learned feature representations}
        & \cell{%
            \tabitem Flexible nonlinear modeling\\
            \tabitem Enhanced feature harmonization\\
            \tabitem Reduced risk of image synthesis artifacts}
        & \cell{%
            \tabitem Reduced interpretability\\
            \tabitem Limited cross-task transferability}
        & {}\\
            
        \midrule 
        \makecell[l]{Traveling-subject\\-based \\(\Cref{sec:ts})}
        & \cell{Paired inter-site differences}
        & \cell{%
            \tabitem Direct removal of site effects\\
            \tabitem Minimal modeling assumptions}
        & \cell{%
            \tabitem Requires traveling subjects\\
            \tabitem Limited scalability}
        & \cell{ 
            \tabitem Traveling subjects}\\
        \bottomrule
        \makecell[l]{}
    \end{tabular}
\end{table*}

\subsection{Image or feature similarity and visual assessment}
\subsubsection{Reference-based evaluation}
Reference-based evaluation is generally regarded as the most direct and interpretable strategy for assessing harmonization performance, as the underlying biological or physical state can be reasonably assumed to remain unchanged. Common forms include traveling subjects and phantoms. Such evaluation is not limited to the retrospective use of traveling-subject datasets described in \Cref{sec:ts}, but also encompasses prospective validation in harmonized data acquisition frameworks (\Cref{sec:acq}).

For image-based harmonization methods, traveling subjects enable direct voxel-wise comparison between harmonized images, allowing the use of quantitative image similarity metrics such as Peak Signal-to-Noise Ratio (PSNR), Structural Similarity Index Measure (SSIM), and Mean Absolute Error (MAE) \citep{RN156,RN73,RN69}. In the case of feature-level harmonization, features extracted after harmonization can be evaluated using paired statistical tests (e.g., paired t-tests), Bland–Altman analysis, intra-class correlation coefficients (ICC), and coefficients of variation (CoV), thereby quantifying cross-site consistency and reliability. Compared with traveling subjects, phantoms provide a more controllable and repeatable imaging object, while their ability to represent true biological variability is inherently limited. Their standardized manufacturing and calibration protocols, such as those established by the National Institute of Standards and Technology (NIST), allow phantom-based validation to be performed across multiple sites or centers without requiring traveling scans. These phantoms have been widely used to assess the consistency of quantitative parameters in harmonized acquisition, particularly in qMRI and diffusion MRI \citep{RN12,RN9,RN8,RN17}.

\subsubsection{Qualitative and visual assessment}
Qualitative visualization strategies for harmonization evaluation can generally be categorized into two classes: feature-level and image-level visualizations. Feature-level visualization focuses on examining the distributions of extracted features across different sites under matched or comparable biological conditions. Typical approaches include visualizing normalized intensity histograms\citep{RN51,RN90}. For harmonized features or learned latent representations, dimensionality reduction techniques such as Principal Component Analysis (PCA) and t-distributed stochastic neighbor embedding (t-SNE) are commonly used to visualize site-related separability before and after harmonization. In addition, some learning-based methods explicitly learn low-dimensional representations during model training, enabling direct visualization of how samples from different sites are geometrically aligned in the representation space following harmonization \citep{RN69,RN72,RN66,RN73}.

In contrast, image-level visualization is primarily adopted in image-based harmonization and harmonized acquisition studies, where evaluation relies on direct visual inspection of contrast consistency and style differences across sites. Such visual assessment remains particularly important in pathology-sensitive scenarios, as it provides one of the few practical means to assess hallucination risk, and to identify spurious structures or anatomically implausible alterations introduced by harmonization procedures \citep{RN74}.

\subsection{Statistical evaluation}
\subsubsection{Reduction of site-related variability}
Although existing studies employ a seemingly diverse set of statistical tests and classification experiments, their underlying objective is largely the same: to determine whether detectable site-related differences remain in imaging features after harmonization. In current practice, this objective is typically addressed through two complementary evaluation strategies.

Statistical testing, including univariate tests and regression-based approaches, is commonly used to assess whether imaging-derived measurements, such as ROI-level summary features, remain significantly associated with site. \citep{RN99,RN95,RN46}. These approaches are based on the assumption that, if harmonization is successful, individual features should no longer exhibit systematic differences across sites. A typical implementation treats site as a categorical factor in ANOVA or linear models to quantify residual mean shifts between sites. Other tests, including Bartlett’s or Levene’s tests, focus instead on variance or scaling differences, and are used to evaluate whether harmonization also corrects site-specific differences in noise magnitude or dispersion. More general distributional differences beyond mean and variance can be assessed using nonparametric tests such as the Kolmogorov-Smirnov test \citep{RN96,RNJMI}.

A second widely adopted strategy is site discriminability testing, in which harmonized images or features are used as inputs to train classifiers (support vector machines or XGBoost) to predict the acquisition site \citep{RN99,RN5,RN3,RN4}. Within this framework, a reduction in site classification accuracy is interpreted as evidence of successful harmonization. More generally, this framework is not limited to site labels and can be extended to other acquisition-related factors, such as field strength, scanner vendor, or imaging protocols. In practice, these factors are often considered components of site-related variability, and site labels are frequently used as proxy variables to implicitly capture such differences. Compared with univariate testing, site discriminability provides a stronger assessment of residual site signatures in high-dimensional feature spaces. However, it is important to note that reduced site predictability does not necessarily imply adequate preservation of biological signals, as excessive harmonization may also remove meaningful biological variability while still achieving low site classification performance.

\subsubsection{Preservation of biological variability}
Compared with evaluating the reduction of site effects, verifying the preservation of biologically meaningful variability is more challenging. Although existing evaluations remain incomplete, prior studies have made several efforts in this direction, including association analyses with biological variables, robustness tests under confounding, and false-positive or permutation-based sanity checks.

Among existing evaluation strategies, biological variable association analysis is the most widely used \citep{RN99,RN5,RN3,RN4}. The core question addressed by this strategy is whether statistical relationships between imaging features and known biological variables (e.g., age, sex, or clinical diagnosis) are preserved after harmonization. In practice, these relationships are typically assessed using generalized linear models (GLM) or ANOVA. Statistical significance is first evaluated to determine whether imaging features remain associated with biological variables after harmonization, commonly reported using z-statistics or p-values. Beyond significance testing, the strength of these associations is further quantified using regression coefficients or other measures of effect size. In addition, some studies further assess biological preservation by examining the proportion of variance in imaging features explained by biological covariates. A representative example is provided by ComBat \citep{RN95}, which evaluates the linear association between average cortical thickness and age before and after harmonization, using changes in $R^2$ as a quantitative indicator of age-related variance preservation.

Confounding between biological variables and site effects poses an additional challenge for MRI harmonization. If not carefully addressed or explicitly modeled, harmonization procedures may inadvertently remove biologically meaningful age-related variability while attempting to eliminate site-related effects. One potential strategy is to explicitly construct datasets with known confounding structures to evaluate harmonization robustness \citep{RN99}. For example, both positively and negatively confounded scenarios can be derived from existing data, or predefined biologically associated and null voxels/ROIs can be used to compare effect sizes before and after harmonization. These analyses assess whether improved sensitivity to true biological effects is achieved without compromising specificity through spurious signal amplification.

Finally, given that harmonization procedures may artificially amplify apparent biological associations even in the absence of true biological differences, some studies further incorporate false positive rate assessments or permutation-based sanity checks \citep{RN5}. The central idea is to repeat association or prediction analyses after randomly permuting biological labels, if harmonized data continue to exhibit statistically significant associations or prediction performance exceeding chance levels, this suggests that the method may have introduced spurious biological signals. Recent study has demonstrated that such strategy constitutes an effective and practical approach for detecting hallucination effects in learning-based methods \citep{RN5}.

\subsection{Performance on downstream tasks}\label{sec:dt}
Downstream tasks provide an additional important dimension for evaluating the harmonization methods in real-world applications. Such evaluations commonly rely on multi-fold cross-validation to fully leverage available data, while also assessing model generalizability under cross-site or cross-dataset validation. Existing studies can be broadly categorized into three types:
\begin{enumerate}
  \item \textbf{Classification}: Representative tasks include binary classification between patient and control groups, as well as multi-class classification of disease subtypes. These tasks are commonly used to assess whether harmonization helps mitigate site-related effects in disease classification. Typical evaluation metrics include the area under the receiver operating characteristic curve (AUC), classification accuracy (ACC), sensitivity (SEN), and specificity (SPE) \citep{RN73,RNMEDIADA} .

  \item \textbf{Segmentation}: Typical tasks include brain tissue segmentation (e.g., gray matter, white matter, and cerebrospinal fluid) as well as lesion segmentation. Such tasks are most frequently conducted in structural MRI (e.g., T1- and T2-weighted imaging) and are primarily used in image-based harmonization. Common quantitative metrics include the Dice Similarity Coefficient (DSC) and the 95th percentile Hausdorff Distance (HD95) \citep{RN74,RN73,RN72,RN69}.

  \item \textbf{Regression}: A representative application is age prediction based on whole-brain signals or imaging-derived features (e.g., cortical thickness), which is used to assess the extent to which harmonization methods preserve continuous biological gradients. Common evaluation metrics include the root mean squared error (RMSE), mean squared error (MSE), and MAE \citep{RN95,RN69,RN61}.
\end{enumerate}

However, it is important to note that improvements in downstream task performance do not necessarily imply biologically faithful or valid harmonization. Performance gains may arise from effective suppression of site-related variability, but may also result from over-correction or the attenuation of biologically meaningful information that is not studied in the specific task.

\subsection{Best practices and checklist}\label{sec:bplist}
Given the differences in data characteristics and methodological objectives across harmonization approaches, we outline a set of general, method-agnostic principles to serve as practical guidance. Harmonization performance should be assessed using multiple complementary criteria, as no single metric is sufficient to fully characterize performance.

\begin{enumerate}
\item \textbf{Explicit assessment of site-effect removal:} For image and feature representations, this can be assessed using site discriminability testing. For feature-level analysis, this can be assessed using statistical tests and low-dimensional visualization.

\item \textbf{Explicit verification of biological signal preservation:} The preservation of biologically meaningful variation, which is often overlooked, should be explicitly validated to avoid overcorrection, for example by biological variable association analysis.

\item \textbf{Report downstream task performance:} Design classification, segmentation, or regression tasks to provide a comprehensive, outcome-level evaluation of method performance.

\item \textbf{Use of reference or ground-truth validation whenever possible:} For acquisition-level methods, prospective validation using traveling subjects or phantoms is preferred. For retrospective studies, available traveling-subject datasets described in \Cref{tab:tab2} can be used.

\item \textbf{Assessment of false-positive and hallucination risks:} Permutation-based sanity checks, combined with downstream tasks and statistical analyses, should be used to evaluate spurious biological associations. Visual assessment is recommended to verify the fidelity of anatomical structures and lesions.
\end{enumerate}

\section{Discussion}\label{sec:cd}
\subsection{Choosing between harmonization strategies}
From a practical standpoint, the choice of harmonization strategy should be guided by data availability, study design, and downstream objectives. Additional considerations include methodological feasibility, such as the availability of open-source implementations, computational requirements, and the degree of reliance on vendor collaboration. Based on these factors, we provide practical recommendations across acquisition-, image-, and feature-level harmonization.

Although acquisition-level harmonization is still in an early stage of development, it represents a promising strategy for prospective multi-center studies based on newly developed sequences or methodologies. In particular, vendor-neutral sequence frameworks such as Pulseq enable rapid multi-site and multi-vendor deployment with minimal reliance on vendor-specific support, thereby bypassing the need for complex and time-consuming platform-specific development and calibration \citep{RN14}. In addition, online reconstruction offers a fast and practical pathway for harmonization in methods that rely on advanced reconstruction techniques, such as qMRI and dMRI. It may also offer advantages in scenarios with substantial site- and operator-dependent variability, including cardiac and fetal imaging \citep{RN23,RNRADAI_PLANES}. However, current acquisition-level approaches remain limited in addressing all sources of data heterogeneity summarized in \Cref{tab:tab1} (e.g., hardware-related factors), and are further constrained by limited flexibility, dependence on vendor support, and insufficient large-scale validation. Therefore, rather than serving as a standalone solution, combining acquisition-level strategies with retrospective harmonization methods may provide a more effective approach to mitigating site effects.

Image-level harmonization is most appropriate when the goal is to produce usable harmonized images, for example, in clinical workflows, expert visual assessment, or downstream analyses that are not yet defined at the time of harmonization. In these scenarios, task-agnostic correction enables broad reuse of the harmonized data, supports subsequent processing steps such as segmentation, classification, lesion visualization, and facilitates long-term data value and public dataset construction \citep{RNSCIDATA_ABCD}. Moreover, image-level approaches offer the potential to mitigate spatially non-uniform or anatomically coupled site effects that cannot be adequately modeled at the feature level. When paired data are available, such as multi-contrast acquisitions or traveling-subject designs, image-level methods can exploit these constraints to better preserve anatomical consistency \citep{RN32,RN33,RN66}. However, in the absence of paired data, image-level harmonization becomes substantially more challenging and typically relies on weaker implicit assumptions, increasing the risk of bias. In particular, in the presence of substantial imbalance in biological covariates across sites, current image-level methods remain limited and may be prone to overcorrection.

Feature-level harmonization remains the dominant choice in current retrospective studies and is expected to continue playing a central role, largely represented by ComBat and its extensions \citep{RN99,RN95}. These methods are particularly well suited for hypothesis-driven studies with clearly defined analysis targets, where downstream tasks and imaging features are fixed and statistical inference is the primary objective. Feature-level approaches are also preferable when sample sizes are limited or site distributions are highly imbalanced, conditions under which image-level learning-based methods are prone to learning site-specific artifacts, introducing reference-site bias, or generating hallucinated effects that are difficult to detect. In addition, feature-level methods allow explicit modeling and control of biological covariates, which is critical when site effects are partially confounded with age, sex, or clinical variables. Within this category, traditional statistical models provide stable and interpretable correction in the absence of paired data, whereas learning-based feature-level methods, such as DeepResBat, offer increased flexibility in modeling nonlinear site effects while still retaining explicit covariate control \citep{RN5}. 

Overall, no single harmonization strategy is universally optimal, and each level entails inherent trade-offs among practicality, flexibility, and biological fidelity. Acquisition-level approaches provide a promising solution when prospective control is feasible, whereas image-level methods offer flexible post hoc correction when harmonized images are required. Feature-level approaches currently represent the most robust and interpretable option for retrospective multi-site studies, particularly under limited or imbalanced data conditions. Careful alignment between study design, data constraints, and evaluation objectives is therefore essential to achieve effective site effect mitigation while preserving meaningful biological variability.

\subsection{Confounding effects and design trade-offs}
A central challenge in multi-site MRI harmonization arises from confounding between site effects and biological covariates, which are often unevenly distributed across sites. In such settings, site-related variability is not independent of biologically meaningful variation, making harmonization fundamentally ill-posed. Existing approaches attempt to mitigate this issue through different modeling strategies. Statistical methods such as ComBat explicitly incorporate biological covariates as fixed effects to preserve their associated variation while removing site-related bias \citep{RN99,RN95}. More recent learning-based methods have introduced additional mechanisms. For example, DeepResBat performs harmonization on residual representations after regressing out covariate effects, thereby reducing the risk of entanglement \citep{RN5}. ImUnity incorporates a biological preservation module by enforcing covariate prediction constraints in the latent space, while deep unlearning methods employ balanced subsets during the deconfounding stage \citep{RN73}. Despite these efforts, severe confounding leads to a fundamental identifiability problem. The degree of identifiability is intrinsically linked to the risk of overcorrection: as identifiability decreases, the ambiguity between site-related and biologically meaningful variation increases, making it increasingly difficult for harmonization methods to distinguish between the two.

This challenge naturally manifests as a key design trade-off in harmonization methods, particularly in deep learning: invariance versus controllability. Methods that enforce site invariance, such as GAN-based approaches, aim to eliminate all site-related differences, but may mistakenly remove biologically meaningful signals when these are correlated with site. Conversely, approaches that emphasize controllability, such as disentanglement-based models, can separate anatomical and style representations and enable flexible manipulation. In this context, implicit strategies allow the model to allocate part of its representational capacity to capture site-specific characteristics, while maintaining a shared latent subspace to encode site-invariant anatomical information, thereby mitigating the risk of overcorrection. However, these methods may also suffer from imperfect disentanglement, leading to bias leakage or unintended alterations of anatomical structure. 

Given this inherent trade-off, harmonization performance should be evaluated within a structured framework that jointly considers site-effect removal and biological signal preservation, as discussed in \Cref{sec:bplist}. Site-effect removal can be assessed through site discriminability tests, while biological preservation can be evaluated by measuring the association between imaging features and relevant covariates. More robust strategies include silver-standard comparison, where a well-balanced dataset (with minimal confounding) is used to estimate reference effect sizes. Harmonization methods can then be applied to artificially confounded datasets, and their ability to recover the reference effects can be quantified \citep{RN99}. In addition, permutation-based sanity checks can be employed to ensure that models do not introduce spurious biological associations while enforcing invariance \citep{RN5}. These considerations emphasize that harmonization should be evaluated as a multi-objective optimization problem, requiring careful evaluation of how well a method navigates the trade-off between invariance and preservation.

\subsection{Harmonization beyond neuroimaging}
Compared with neuroimaging, MRI harmonization studies in other anatomical regions (e.g., cardiac and abdominal imaging) remain relatively limited and are predominantly application-driven. Existing approaches mainly focus on feature-level harmonization, including first- and higher-order radiomic features as well as deep learning-derived representations, with ComBat and its variants widely adopted to mitigate inter-site variability. Prior studies have demonstrated that such strategies can improve feature robustness in cardiac MRI, facilitate cross-cohort integration in large-scale abdominal imaging studies, and enhance downstream tasks such as tissue classification and prognosis prediction  \citep{RNb001,RNb002,RNb003,RNb004}. However, methodological developments tailored to these anatomical regions remain scarce. Although existing frameworks can be applied to some extent, they often fail to account for region-specific factors, such as respiratory motion, cardiac dynamics, and more heterogeneous morphology and tissue composition, highlighting the need for more tailored harmonization strategies.

\subsection{Clinical translation and deployment challenges}
While harmonization methods are currently most widely used in research settings for multi-center and large-scale studies, there are additional considerations for clinical deployment. Regulatory approval can be a major barrier in clinical deployment of image harmonization methods, especially those based on machine learning \citep{RNRAD_IHEAI}. Regulators such as the FDA (Food and Drug Administration) and EMA (European Medicines Agency) require clear evidence that a harmonization method does not alter clinically relevant information or introduce systematic bias that could affect diagnosis, longitudinal follow-up, or treatment decisions. Demonstrating robust performance across platforms and software versions may require extensive validation. Both FDA\footnote{https://www.fda.gov/regulatory-information/search-fda-guidance-documents/marketing-submission-recommendations-predetermined-change-control-plan-artificial-intelligence} and EMA\footnote{https://www.ema.europa.eu/en/news/reflection-paper-use-artificial-intelligence-lifecycle-medicines} have provided new rules and guidelines for continuous monitoring of methods that may change over time, such as those based on continuous learning and federated learning.

Widespread clinical deployment of harmonization methods may also require vendor support. While developers or users (e.g., hospitals) can potentially maintain custom software, workflows, and ensure security and compliance on their own, image harmonization may be best achieved if it is integrated with vendor reconstruction pipelines. This may not be possible without vendor support or involvement.

\subsection{Future Directions}
With increasing efforts to harmonize the MRI workflow across the full acquisition-to-analysis pipeline, and driven in particular by advances in deep learning, the field is poised to benefit from several emerging opportunities :

\begin{itemize}
  \item \textbf{Harmonized acquisition and reconstruction:}
  Previous research has primarily focused on retrospective harmonization, whereas harmonized acquisition and reconstruction strategies have received comparatively less attention \citep{RN39,RN42,RN43}. As an approach that minimizes variability at the source, this strategy is undoubtedly one of the promising directions for future development. Further progress will depend on simplifying implementation and promoting the sharing of standardized acquisition and reconstruction workflows. When combined with retrospective harmonization approaches, these strategies have the potential to achieve more effective and robust harmonization across sites, while jointly optimizing the overall MRI acquisition pipeline in both clinical and research settings.

  \item \textbf{Establishing standardized validation benchmarks:}
  There has long been an urgent need to establish unified validation benchmarks that are quantitative and directly comparable across studies. The public release of large-scale, multimodal traveling-subject datasets, such as ON-Harmony \citep{RN118}, now provides a critical opportunity to advance this effort. Such benchmarks are particularly important for addressing the current landscape of harmonization research, especially the rapid proliferation of deep learning-based methods, for which evaluation protocols remain heterogeneous. At the same time, existing evaluation frameworks themselves remain limited and call for further methodological innovation, including the quantitative assessment of image-level hallucinations, as well as the evaluation of uncertainty and newly introduced biases associated with learning-based approaches.

  \item \textbf{Integrating statistical and deep learning approaches:}
  While deep learning offers powerful nonlinear modeling capabilities at both the image and feature levels, existing evidence suggests that such models may inadvertently suppress biologically meaningful signals, even when biological covariates are explicitly modeled \citep{RN5}. DeepResBat \citep{RN5} provides a representative example at the feature level, demonstrating that deep learning-based residual modeling can improve harmonization performance while still requiring careful control to avoid overcorrection of biological variability. In contrast, comparable image-level harmonization approaches that explicitly address this trade-off remain largely unexplored. From a methodological perspective, statistical approaches tend to be more robust in small-sample regimes or under severe confounding, whereas deep learning is capable of modeling complex spatial and structural patterns in images. Integrating these complementary strengths may therefore represent a promising direction for developing harmonization methods that are both flexible and biologically faithful.

  \item \textbf{Data privacy and security:}  
  Traditional harmonization methods typically rely on centralized data processing, which is often impractical given the sensitive nature of medical images and cross-institutional data-sharing constraints. Federated learning offers an alternative by enabling local model training with only parameter sharing, thereby, in principle, alleviating the need for centralized harmonization \citep{RN115,RN110}. However, adapting existing harmonization strategies to federated settings remains challenging, as many methods assume joint access to source and target domains or paired training data, conditions that are rarely met in distributed multi-site studies \citep{RN110}. Beyond federated learning, an emerging complementary direction is to explicitly model and simulate unwanted sources of variability during training, as exemplified by synthetic data-driven approaches \citep{RN105,RN106,RN108,RN103,RN113}. By exposing models to controlled, diverse forms of site-related bias, such strategies aim to improve robustness and reduce data-sharing constraints, offering a promising alternative for privacy-preserving harmonization.

  \item \textbf{Foundation model:}  
  Large-scale pretraining on diverse datasets enables high-capacity encoders to learn domain-invariant representations, inducing a degree of implicit harmonization \citep{RNf3,RNf4}. As demonstrated by models such as BME-X \citep{RN102} and BrainIAC \citep{RNf2}, this process enhances robustness to site-specific biases, including variations in contrast and intensity. Nevertheless, traditional harmonization methods remain indispensable due to their interpretability and practicality. Therefore, future work may focus on integrating foundation models with harmonization strategies. For example, explicit harmonization in the pretrained latent space may provide additional stability to learned representations and improve robustness when transferring to resource-limited or previously unseen sites.
\end{itemize}

In summary, harmonization as a preprocessing step for large-scale MR image analysis continues to offer substantial opportunities and plays an important role that is difficult to replace. At the same time, as these methods are increasingly applied and further developed, their limitations and potential risks should be carefully acknowledged. Method selection should therefore be guided by specific analytical goals and supported by rigorous and comprehensive validation, to ensure appropriate and responsible use of harmonization in practice.

\section{Conclusion}\label{sec:co}
This survey reviews recent advances in MRI harmonization across the full imaging pipeline, spanning harmonized image acquisition, retrospective image-level and feature-level methods, evaluation strategies, and publicly available traveling-subject datasets. Nevertheless, although substantial progress has been made in reducing site effects, solid evidence for biological signal preservation and standardized validation benchmarks are still lacking. Accordingly, this survey provides a comprehensive reference for the field and highlights key methodological gaps and future directions toward developing reliable, biologically informed harmonization frameworks for multi-site MRI studies.

\vspace{1em}
\textbf{CRediT authorship contribution statement}

\textbf{Qinqin Yang}: Writing - review \& editing, Writing - original draft, Visualization, Resources, Methodology, Investigation, Conceptualization. 
\textbf{Firoozeh Shomal-Zadeh}: Writing - review \& editing, Writing - original draft, Visualization, Investigation. 
\textbf{Ali Gholipour}: Supervision, Conceptualization, Writing - review \& editing, Validation, Funding acquisition.

\vspace{1em}
\textbf{Declaration of competing interest}

The authors declare that they have no known competing financial interests or personal relationships that could have appeared to influence the work reported in this paper.

\vspace{1em}
\textbf{Acknowledgments}

This research was supported in part by the National Institute of Health (NIH) under award numbers R01EB031849, R01EB032366, and R01HD109395. The content of this publication is solely the responsibility of the authors and does not necessarily represent the official views of the NIH.

\vspace{1em}
\textbf{Data availability}

No data was used for the research described in the article.

\bibliographystyle{model2-names.bst}
\biboptions{authoryear}
\bibliography{refs}

\end{document}